\documentclass[12pt,preprint]{aastex}

\usepackage{ulem}

\newcommand{\ltsimeq}{\raisebox{-0.6ex}{$\,\stackrel
        {\raisebox{-.2ex}{$\textstyle <$}}{\sim}\,$}}

\shorttitle{Dust Properties of 73P}
\shortauthors{Harker et al.}

\begin{document}


\title{MID-INFRARED SPECTROPHOTOMETRIC OBSERVATIONS OF FRAGMENTS B AND C 
OF COMET 73P/SCHWASSMANN-WACHMANN 3} 



\author{
DAVID E.\ HARKER\altaffilmark{1},
CHARLES E. WOODWARD\altaffilmark{2},\\
MICHAEL S.\ KELLEY\altaffilmark{3},
MICHAEL L.\ SITKO\altaffilmark{4},
DIANE H.\ WOODEN\altaffilmark{5},\\
DAVID K.\ LYNCH\altaffilmark{6},
RAY W.\ RUSSELL\altaffilmark{7},
}

\altaffiltext{1}{Center for Astrophysics and Space Sciences, University of 
California, San Diego, 9500 Gilman Drive, La Jolla, CA 92093-0424,\ 
{\it{dharker@ucsd.edu}} } 

\altaffiltext{2}{Department of Astronomy, School of Physics and
Astronomy, 116 Church Street, S.~E., University of Minnesota,
Minneapolis, MN 55455,\ {\it{chelsea@astro.umn.edu}} }

\altaffiltext{3}{The University of Maryland, Dept. of Astronomy, College
Park, MD 20742-2421}

\altaffiltext{4}{Space Science Institute, Boulder, CO 80301}
 
\altaffiltext{5}{NASA Ames Research Center, Space Science Division, 
MS~245-1, Moffett Field, CA 94035-1000, \ {\it{diane.h.wooden@nasa.gov}} } 

\altaffiltext{6}{Thule Scientific, P.O. Box 953, Topanga, CA 90290}

\altaffiltext{7}{The Aerospace Corporation, Los Angeles, CA 90009}

\begin{abstract}

We present mid-infrared spectra and images from the GEMINI-N ($+$Michelle) 
observations of fragments SW3-[B] and SW3-[C] of the ecliptic (Jupiter Family) 
comet 73P/Schwassmann-Wachmann~3 
pre-perihelion.  We observed fragment B soon after an outburst event 
(between 2006 April 16 -- 26~UT) and detected crystalline silicates. The 
mineralogy of both fragments was dominated by amorphous carbon and 
amorphous pyroxene.  The grain size distribution (assuming a Hanner
modified power-law) for fragment SW3-[B] has a peak grain radius of 
$a_{p} \sim 0.5$~\micron, and for fragment SW3-[C], $a_{p} \sim 0.3$~\micron; 
both values larger than the peak grain radius of the size 
distribution for the dust ejected from ecliptic comet 9P/Tempel~1 during the 
Deep Impact event ($a_{p} = 0.2$~\micron).   The silicate-to-carbon ratio 
and the silicate crystalline mass fraction for the submicron to micron-size 
portion of the grain size distribution on the nucleus of fragment 
SW3-[B] was $1.341^{+0.250}_{-0.253}$ and $0.335^{+0.089}_{-0.112}$, respectively, 
while on the nucleus of fragment SW3-[C] was $0.671^{+0.076}_{-0.076}$ and
$0.257^{+0.039}_{-0.043}$, respectively.
The similarity in mineralogy and grain properties between the two fragments 
implies that 73P/Schwassmann-Wachmann~3 is homogeneous in composition.
The slight differences in grain size distribution and silicate-to-carbon 
ratio between the two fragments likely arises because SW3-[B] was actively fragmenting
throughout its passage while the activity in SW3-[C] was primarily driven by
jets.
The lack of diverse mineralogy in the 
fragments SW3-[B] and SW3-[C] of 73P/Schwassmann-Wachmann~3 along with the 
relatively larger peak in the coma grain size distribution suggests the 
parent body of this comet may have formed in a region of 
the solar nebula with different environmental properties than the natal 
sites where comet C/1995~O1 (Hale-Bopp) and 9P/Tempel~1 nuclei 
aggregated. 
  
\end{abstract}

\keywords{comets: general: comets: individual (73P/Schwassmann-Wachmann~3)
:dust: modeling}

\section{INTRODUCTION}
\label{sec:intro}

Comet nuclei contain some of the least thermally- and least 
aqueously-altered materials extant in the solar system and are highly 
porous agglomerates of ice and dust grains, perhaps with highly stratified, 
inhomogeneous layers of varied density, porosity, and composition 
\citep{hark07,belton06} formed by layered accretion of icy planetesimals. 
The latter may or may not have aggregated from specific radial zones in the 
protoplanetary disk to form the nascent nuclei of comets, ultimately giving 
rise to the two comet families, ecliptic comets (ECs; including Jupiter 
Family comets and active Centaurs) and nearly isotropic comets (NICs; 
including long-period Oort cloud and Halley-type comets). 

Recent dynamical models by the `Nice group' indicate that Jupiter
ejected bodies from the solar system's gravitational potential and
into the Oort Cloud to become NICs.  Comet nuclei that formed in the 
trans-Neptune region, {\it in situ} in the Kuiper Belt (KB), and the 
possible few ejected by Jupiter into the KB, survived dynamically to become 
ECs \citep{morb04,whitman06}. 
As collisions are frequent in the Kuiper Belt \citep{pan05}, 
Jupiter-family comets (ECs) could be chips off of larger Kupier Belt
Objects (KBOs).  
Therefore, the structure of EC nuclei may be very different from NICs.  
On the other hand, Oort cloud (NIC) comets (i.e., long-period) were 
gravitationally scattered to the Oort cloud \citep{oort51} where they 
suffered no further collisions \citep{stern03}. Once perturbed into the 
inner solar system, NICs have short life times compared to EC comets
\citep{ahearn85}. ECs have notably less activity than NICs
\citep{ahearn85, meech04}. Lower activity and lower dust production
rates result in less thermal dust emission making dust properties of
ECs relatively unknown compared to brighter NICs 
\citep{woo07, sitko04, hanner94}.

The strongest tests for new dynamical models of solar system formation 
processes are the compositions of comets, and in particular, the diversity 
of the gas-to-dust ratio and dust composition within any given comet's 
coma coupled with variations in the latter physical characteristics (among 
others) between comets within a dynamical 
family and between dynamical families. Notably, solar nebula models 
for thermal processing and radial transport \citep{boss04,cuzzi93} of 
grains are constrained by the crystalline silicate mass fraction
deduced for comets \citep{woo07,woo05}, as well as the (controversial) 
presence of aqueous altered grain components.  
Amorphous silicates in comets are thought 
to be presolar ISM silicate grains \citep{wooden02,hanbrad04} 
because of their high abundance in the ISM \citep{lidraine01} and 
ubiquitous presence as glassy spherules (GEMS) in cometary 
interplanetary dust particles \citep{brad94,brad99,hanbrad04}.
On the other hand, crystalline silicates are relatively rare 
(\ltsimeq 2.2 -- 5\%) in the 
ISM \citep{kemper04,kemper05,lidraine01} yet are abundant in many comets.  
Crystalline silicates in comets are either high-temperature ($\approx$1450~K) 
condensates \citep{grossman72}, or amorphous silicates annealed into crystals 
at $\sim$1000~K \citep{hallen98, woo05, woo07} in the inner zones of the 
early (\ltsimeq 0.3 Myr-old) solar nebula \citep{dbm02}, or annealed in 
nebular shocks at 5-10~AU (\ltsimeq 1 Myr-old) \citep{harkdes02}.
In models of turbulent mixing, the early formation of crystals in the 
hot inner zones of the solar nebula are concurrent with rapid disk 
spreading (0.3~Myr) suggests a uniform 
crystalline-to-amorphous silicate ratio between comets \citep{dbm02}.
In contrast, viscous dissipation models may yield a 
radial gradient in the crystalline silicate mass fraction that spreads to 
only 20~AU in 1~Myr \citep{woo07, ciesla07}, making it harder to 
explain crystals in comets that formed in the trans-Neptune zone.  
Strong constraints for nebula models include the 
high crystalline silicate mass fraction ($\sim$25\%--50\%) deduced from 
modeling NIC C/1995 O1 (Hale-Bopp) \citep{hark04},  
NIC C/2001 Q4 (NEAT) \citep{harker05,woo04}, the ejecta from the EC 
9P/Tempel~1 during {\it Deep Impact} \citep{harker05,hark07}, and the EC 
78P/Gehrels~2 \citep{watanabe05}.  \textit{Stardust} samples from EC 
81P/Wild~2 contain abundant crystalline silicates \citep{ebel06} and even 
higher temperature calcium-aluminum inclusion (CAI)-type minerals.  In contrast, 
other ECs have a low ($\ll$50\%) crystalline silicate mass fractions: pre-impact 
9P/Tempel~1 \citep{lisse06}, 103P/Hartley~2 \citep{crovisier00, crov99}, and 
29P/Schwassmann-Wachmann~1 \citep{stansberry04}.  
Thus, assessing the composition of cometary nuclei is critical to understanding 
the origins of comets \citep{ehrenf04, hanbrad04, crovis07} and the formation of 
the solar system. 

Contrasts between EC and NIC family grain properties often rely
on comparisons between ground-based remote-sensing studies 
and the \textit{in situ} sampling of coma dust in comets such as 
1P/Halley, 81P/Wild~2 or the 
9P/Tempel~1 (henceforth 9P) Deep Impact ejecta analysis.
Frequently, the 10~\micron{} spectra of short-period ECs
exhibit a broad and weak 10~\micron{} feature, of order 10-20\%
above the 8 to 13~\micron{} blackbody continuum \citep{han_zol10}
and little evidence for the presence of crystalline
silicates -- a result of
space weathering, thermal processing, and de-volatilization arising
from frequent perihelion passage, whereas in NIC comets such emission
is common \citep{mskdhw09}.
Surprisingly, the Deep Impact event excavated sub-surface materials 
containing a significant population of small grains, including 
crystalline silicate species, unlike the grain population dominant in the
quiescent coma of the EC 9P.  The observed 10 and 20~\micron{}
spectra and composite thermal spectral energy distribution (SED)
models of the 9P dust ejecta were similar in characteristics to those
deduced for the coma grains in comet C/1995~O1 (Hale-Bopp), a NIC. 
\citet{hark07}, \citet{sugita05}, and
\citet{lisse06} among others conjectured that the interior of 9P may
be comprised of less processed, more primitive, or pristine grain
material.  Therefore, inference between comet families drawn from the
study of coma dust is a challenge.  
The assertion that ECs are compositionally distinct from NICs as a class
is no longer valid based on the Deep Impact experiment 
\citep{han_zol10}.

In its 1995 apparition, the Jupiter-family comet
73P/Schwassmann-Wachmann~3 (henceforth SW3) exhibited a sudden
increase in brightness \citep{crovis96} near perihelion breaking into
three bright fragments and several fainter components
\citep{scotti96}.  The observing geometry of SW3's next perihelion
  passage (orbital period $T = 5.36$~yr) was not very favorable for
  Earth-based observers \citep{sekanina05}. During SW3's 2006 return
  to perihelion $\approx 66$ fragments were detected \citep{weaver08},
  many of which were remnants of the splitting event of 1995
  \citep{reach09}.  Fragments SW3-[B] and [G] continued to split, and
  swarms of mini-comets were discovered in \textit{Hubble Space Telescope} 
images of the tails of these fragments \citep{weaver08}. 
 Fragmentation of this comet's nucleus,
a natural analogy to the Deep Impact mission, provided a 
rare opportunity to observe potentially pristine material from
the progenitor interior \citep{dellorusso07}.

Here we present mid-infrared narrowband imagery and spectroscopy
(\S\ref{sec:observation}) and coma dust mineralogy derived from
thermal grain models (\S~\ref{sec:model}) of the two largest surviving
fragments, SW3-[B] and SW3-[C] during the 2006 apparition, discussing
comparison (\S\ref{sec:results}) between grain properties of this
comet (one of the most active known ECs) to the putatively more
primitive dust populating the coma of NIC comets such as C/1995~O1
(Hale-Bopp). Our conclusions are summarized in \S\ref{sec:concl}.


\section{OBSERVATIONS}
\label{sec:observation}

Mid-infrared (IR) 10~\micron\ spectra and 10 and 20~\micron\ images of
fragments SW3-[B] and SW3-[C] of comet SW3 pre-perihelion (perihelion 2006
June 08~UT) on 2006 April 29.5~UT (fragment SW3-[B]) and 2006 April
30.5~UT (fragment SW3-[C]) were obtained using the Michelle imaging
spectrograph on the 8-m Fredrick C.\ Gillett telescope (Gemini-N)
on Mauna Kea, Hawaii conducted under program GN-2006A-DD-1.  In
addition, we obtained 20~\micron\ spectra of fragment SW3-[B].  The
observational summary is shown in Table~\ref{tab:obssum}.

Michelle has several mid-IR broad and narrow band filters 
\citep{alistair97}.  For our observations, we used two narrow band filters: 
1)~the Si-5 filter with a central wavelength of 11.6~\micron\ and width of 
9.5\%; and 2)~the Qa filter with a central wavelength of 18.1~\micron\ and 
width of 10.7\%.  Common mid-IR observing techniques 
\citep[e.g.,][]{joyce92}, chopping the secondary ($\pm 15$\arcsec{} at a 
$45^{\circ}$ angle in the NW-SE direction) and nodding the telescope 
($\pm 15$\arcsec\ in the NW-SE direction) to reduce the thermal noise from 
the background and avoid contamination from the coma extending to the south 
west for both comets, were utilized. The brightness of the coma in the 
NW-SE direction for both fragments drops to the value of the background at 
11.2~\micron{} ($40.12$~mJy arcsec$^{-2}$) at 6\arcsec{} away from the 
centralized core.  A final image is produced by subtracting the two nod 
pairs.  

Michelle uses a grating to produce spectra with wavelength coverage of 
$\approx 7.8-13.2$~\micron\ and $\approx 17.0-21.0$~\micron\ within the 
10~\micron\ and 20~\micron\ ground-based atmospheric windows, 
respectively.  A 0.6\arcsec{} wide slit was used for all of our spectral 
observations, providing a resolution of $R \sim 200$ and $R \sim 110$ at 
10~\micron\ and 20~\micron, respectively.  We chopped on array at the 
maximum allowable chop throw of 15\arcsec{} along the slit
direction, perpendicular to the dispersion axis of Michelle.  


Extraction of the comet and standard star spectra from the
two-dimensional images used common mid-IR data reduction techniques
\citep{joyce92}: 1)~two nod pairs are subtracted to produce a
background subtracted image which contains both a positive and
negative spectrum; 2)~two extractions (spatial dimension of
1.0\arcsec{} yielding an effective aperture diameter of 0.87\arcsec)
are made to produce each spectrum: one is the source spectrum, and the
other is extracted 6\arcsec{} to the north of the source spectrum 
for the purpose of calculating the residual background; 3)~all of the
counts at each pixel (i.e., wavelength) inside the source aperture
are summed; 4)~the sky extraction is used to calculate a median
residual background value and variance, which is used to
calculate the photometric error for the object spectrum; and
5)~the median sky value is subtracted from the object spectrum.
Nominally, a residual sky spectrum to the south of the source spectrum is
also used to calculate the background value and variance.  However,
in this case, for both fragments, there is emission from the extended
coma in this southern direction.  Therefore, we have relied only on
the northern sky extraction. Using this technique does not affect the
overall value of the sky, but simply does not enable better constraint
of the uncertainties in the sky variances. This uncertainty is
reflected in the larger 1-sigma error in the final spectrum.  
 We increased the signal-to-noise of the data with a
 3-point statistically weighted average boxcar function.  This
operation decreased the spectral resolution to $R = 129$ at
10~\micron. The spectral resolution is sufficient to identify mineral
resonances that may exist \citep{hark02,woo99}. In 100.8~s of
on-source integration time, a given flux point in the final spectra
has a root-mean-square (RMS) error of $2.5 \times 10^{-19}$~W
cm$^{-2}$ $\mu$m$^{-1}$.

Cohen standards \citep{cohe96} which provide superior IR spectrophotometric
calibration, and ATRAN atmospheric models
\citep{lord93} are used to flux calibrate the spectra and correct for
atmospheric extinction.  The object spectrum is multiplied by the
Cohen stellar template, divided by the extracted
instrument spectrum of the standard, and finally multiplied by the
ratio of the atmospheric transmission spectra calculated for the
airmass of the comet and standard star.  Photometric standards were 
also used to assess the seeing and derive estimates of the point-spread 
function (PSF). The average PSF FWHM was 0.4$^{\prime\prime}$ at 11.6~\micron\ and
0.5$^{\prime\prime}$ at 18.1~\micron.

\begin{figure}
\epsscale{1.1}
\plottwo{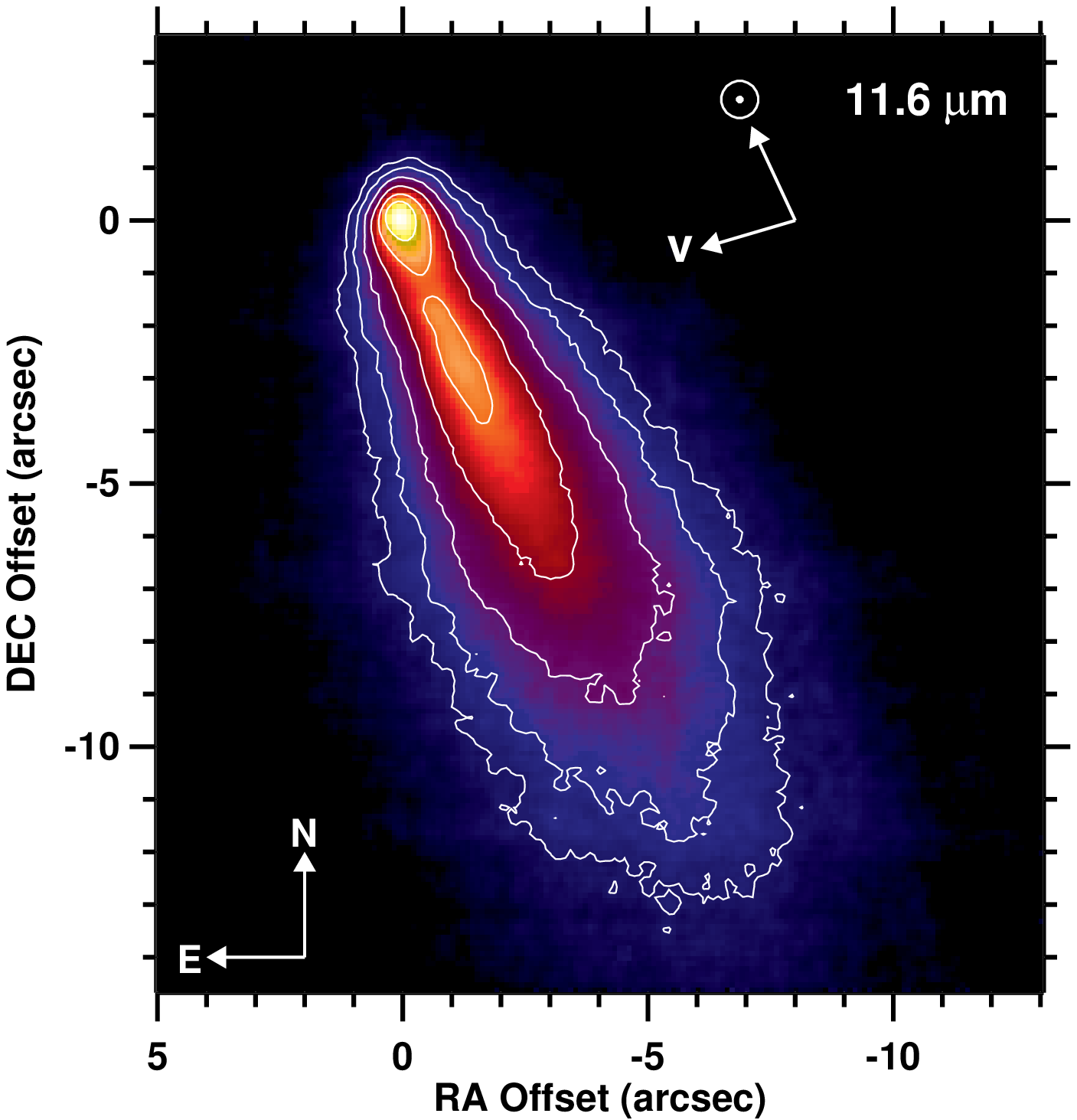}{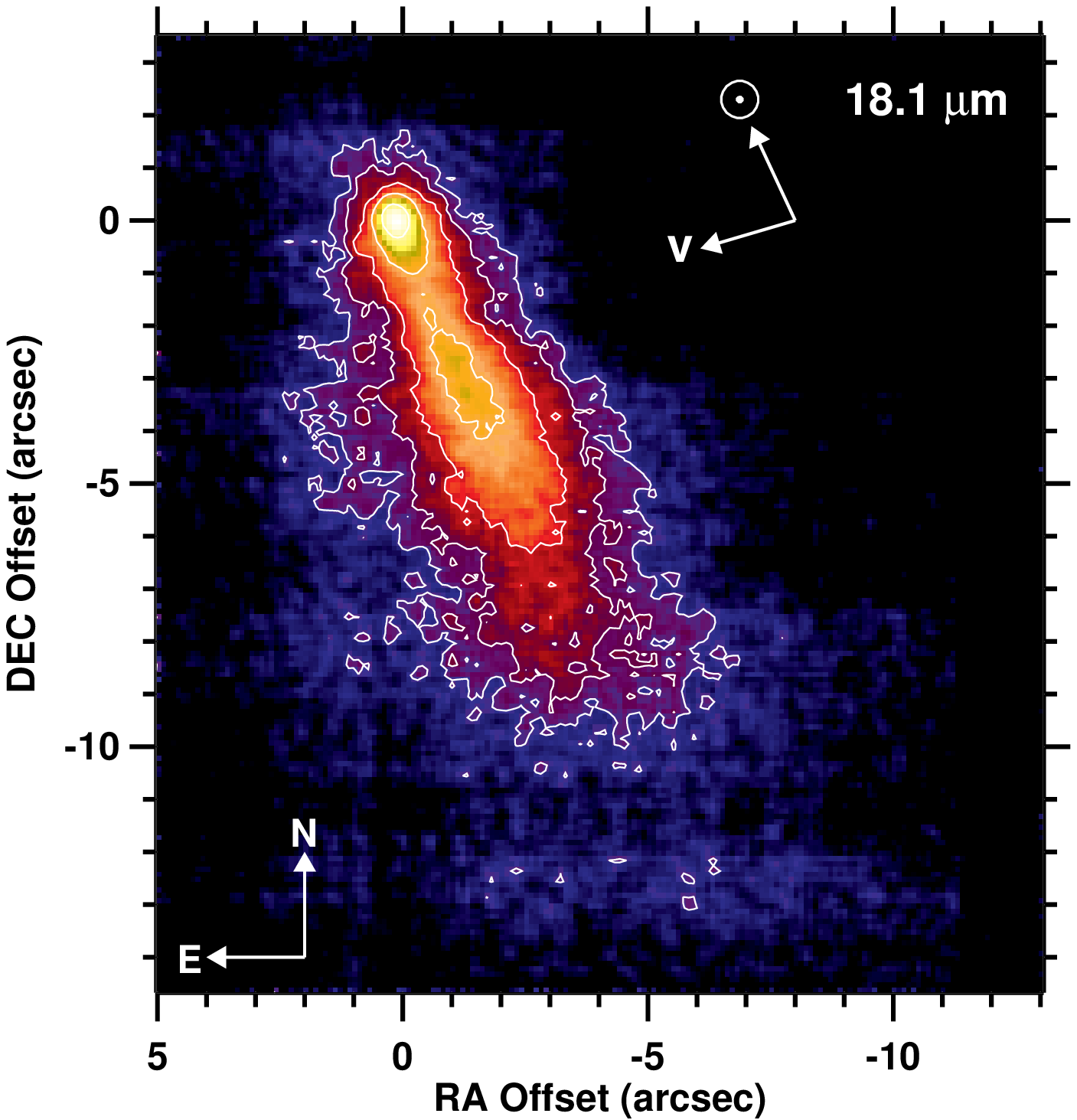}
\plottwo{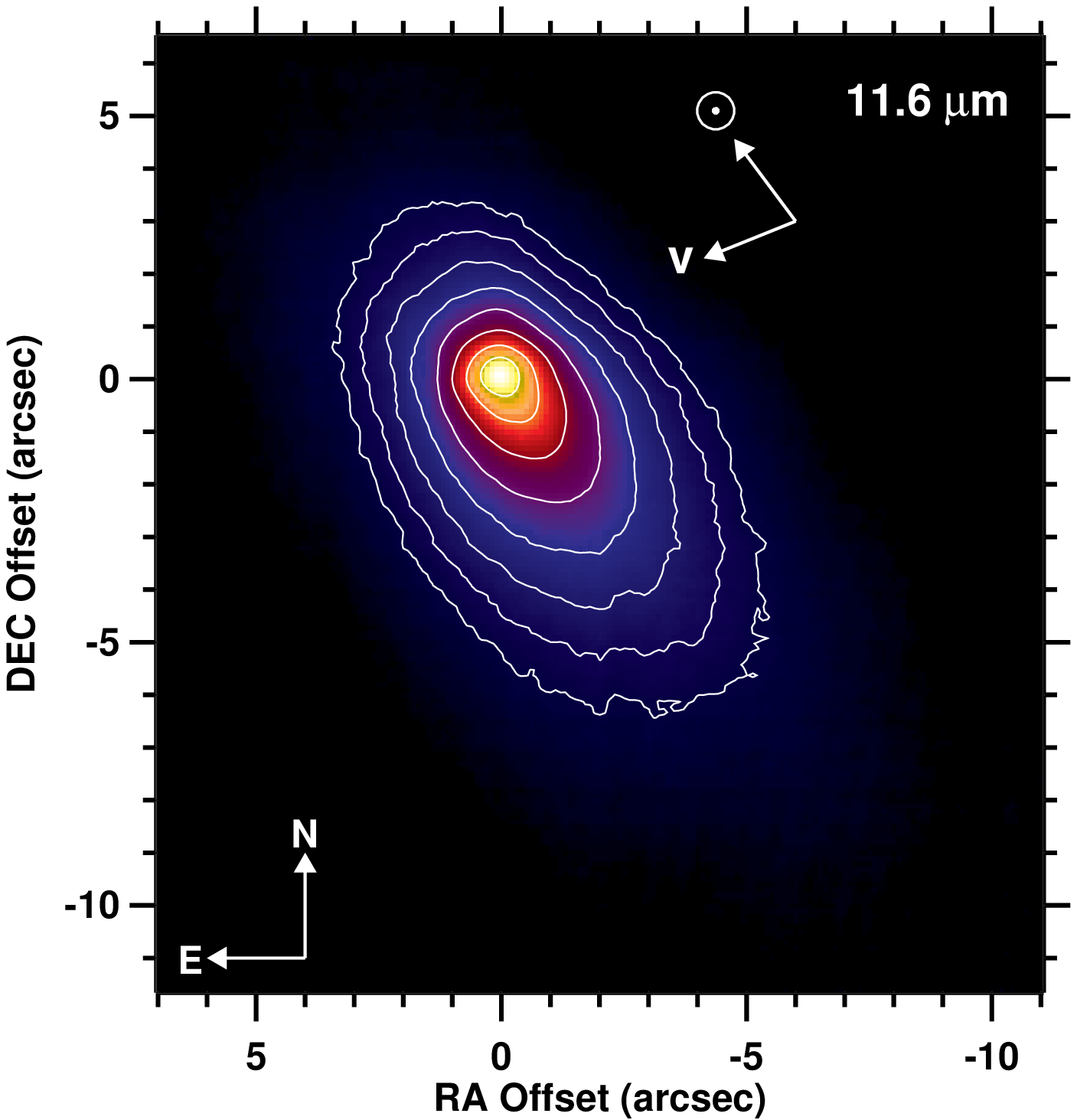}{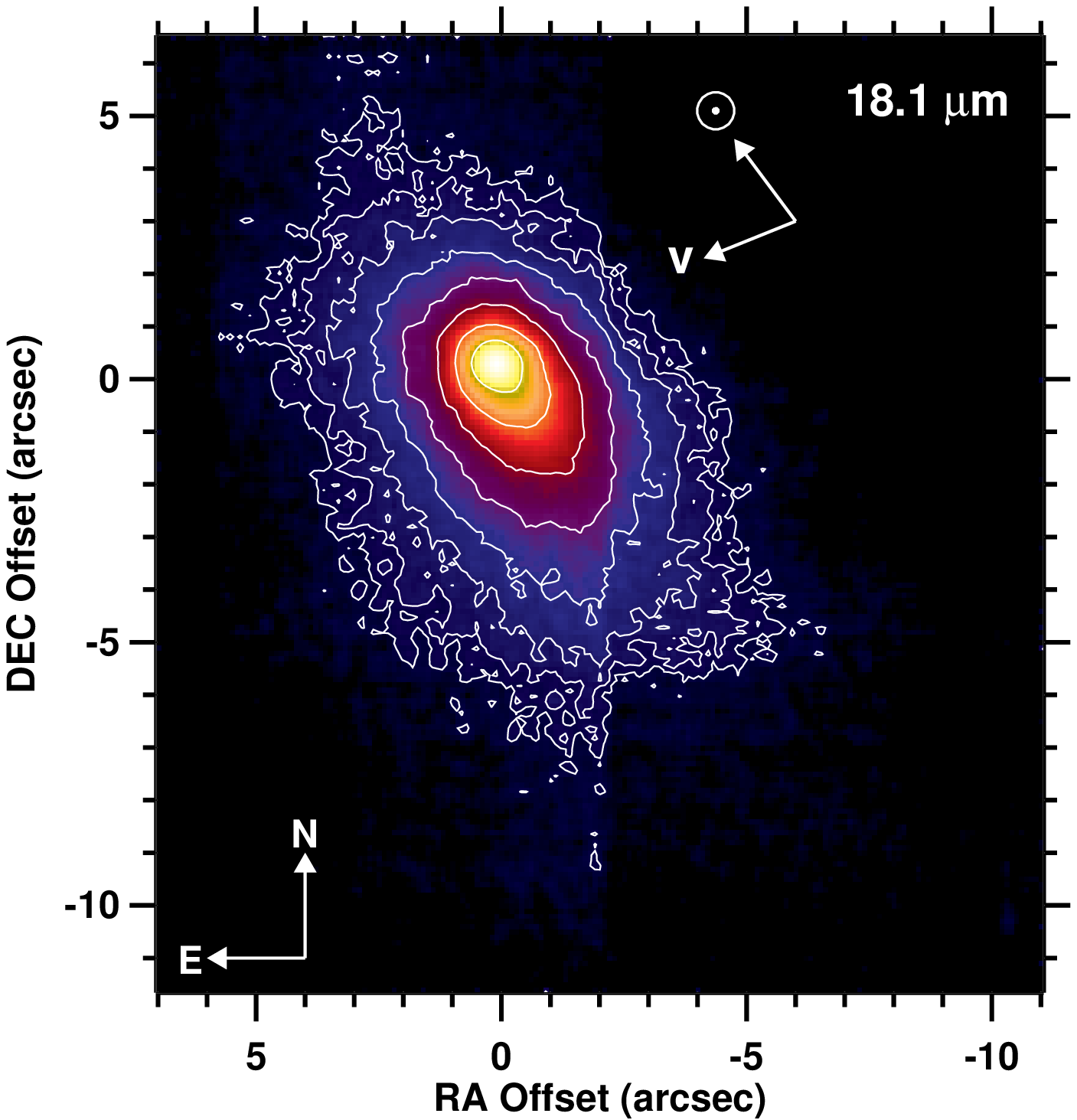}
\caption{Images of comet 73P/Schwassmann-Wachmann~3 fragment SW3-[B] 
({\it top row}) and fragment SW3-[C] ({\it bottom row})
at 11.6~\micron\ ({\it left column}) and 18.1~\micron\ ({\it right column}).
The peak isophotes in the two panels for SW3-[B] represents 17.4~mJy, and for
SW3-[C] represents 76.6~mJy.  The ratio between isophotes in all panels
is $\sqrt{2}$.  The orientation of the frame on the sky, the direction of 
the Sun, and the direction of the velocity vector are indicated by arrows.
The spatial scale of the panels for SW3-[B] is $1828\times1828$~km and for
SW3-[C] is $1697\times1697$~km.
\label{fig:73pb_im} }
\end{figure}

Figure~\ref{fig:73pb_im} shows the 11.6 and 18.1~\micron\ images of 
fragments SW3-[B] and SW3-[C], respectively.  The 3-sigma detection limit of the
coma of SW3-[B] extends $\sim 15$\arcsec{} from the central nucleus region towards 
the south western direction at a position angle (PA) $\simeq 205^{\circ}$ in both images.  
Notably, there is a region of enhanced flux (32\% above the expected flux
of a standard $1/\rho$ decline in coma surface brightness, where 
$\rho$ is the projected distance from the nucleus) approximately 
3\arcsec\ away from the centralized condensed region that is observed at 
both wavelengths.  The coma of SW3-[C] is fairly condensed (0.6\arcsec{}
at 11.6~\micron), the 3-sigma detection limit of the coma extending only about 
$\sim 8$\arcsec{} at PA$\simeq 219^{\circ}$.  We note that more sensitive 
{\it Spitzer} MIPS24 observations of 73P, obtained less than a week after 
our Michelle observations (2006 May 4 -- 5~UT), showed that the surface 
brightness of the coma of fragment SW3-[B] extended some 4$^{\prime}$ while 
that of SW3-[C] extended $\simeq 5^{\prime}$ \citep{reach09}.
Isophote-contour maps derived from our SW3-[C] 
image do not reveal any enhanced regions of flux density, in contrast to 
the surface brightness morphology evident in the coma of fragment SW3-[B]. 

\begin{figure}
\epsscale{1.1}
\plottwo{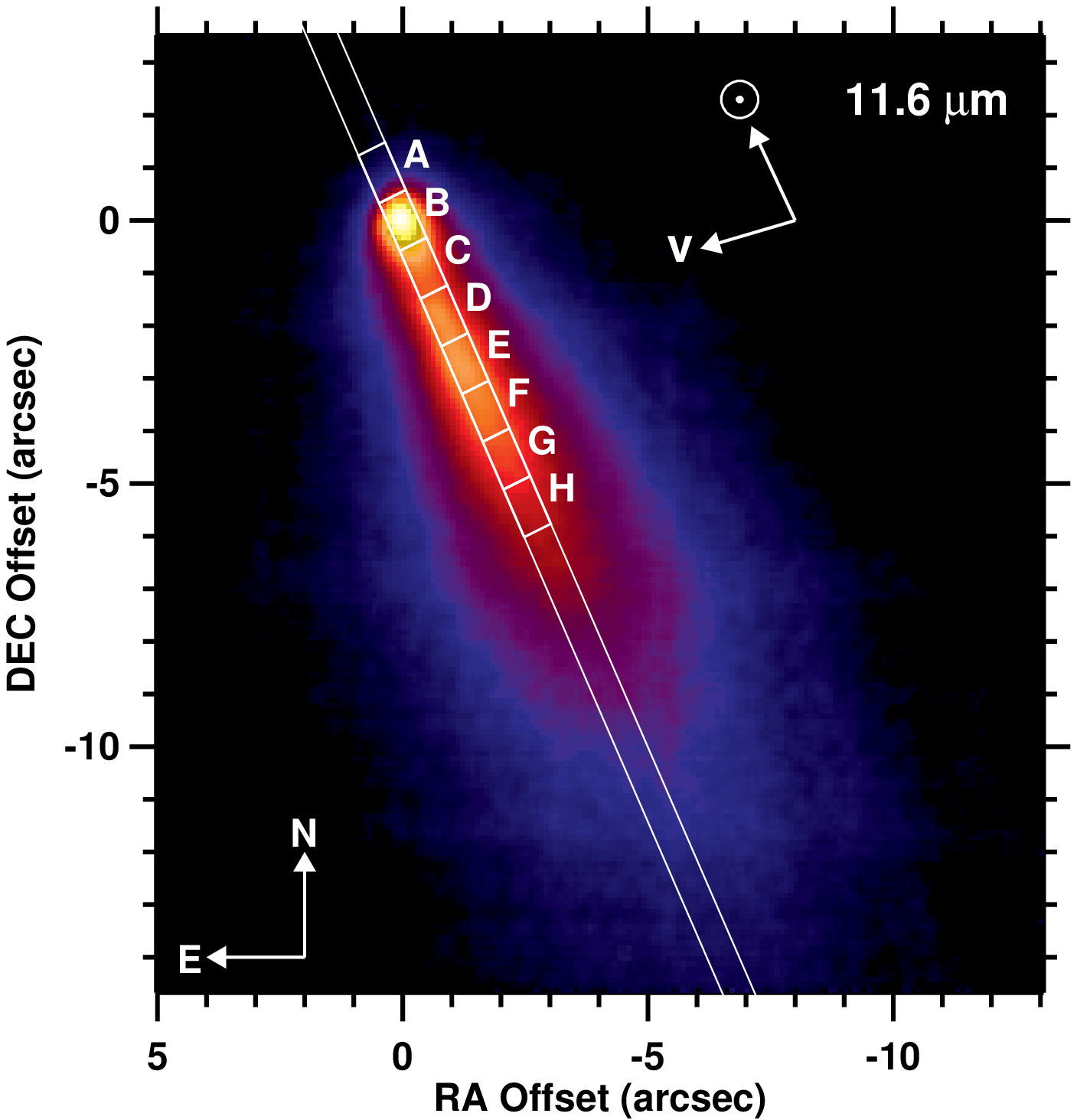}{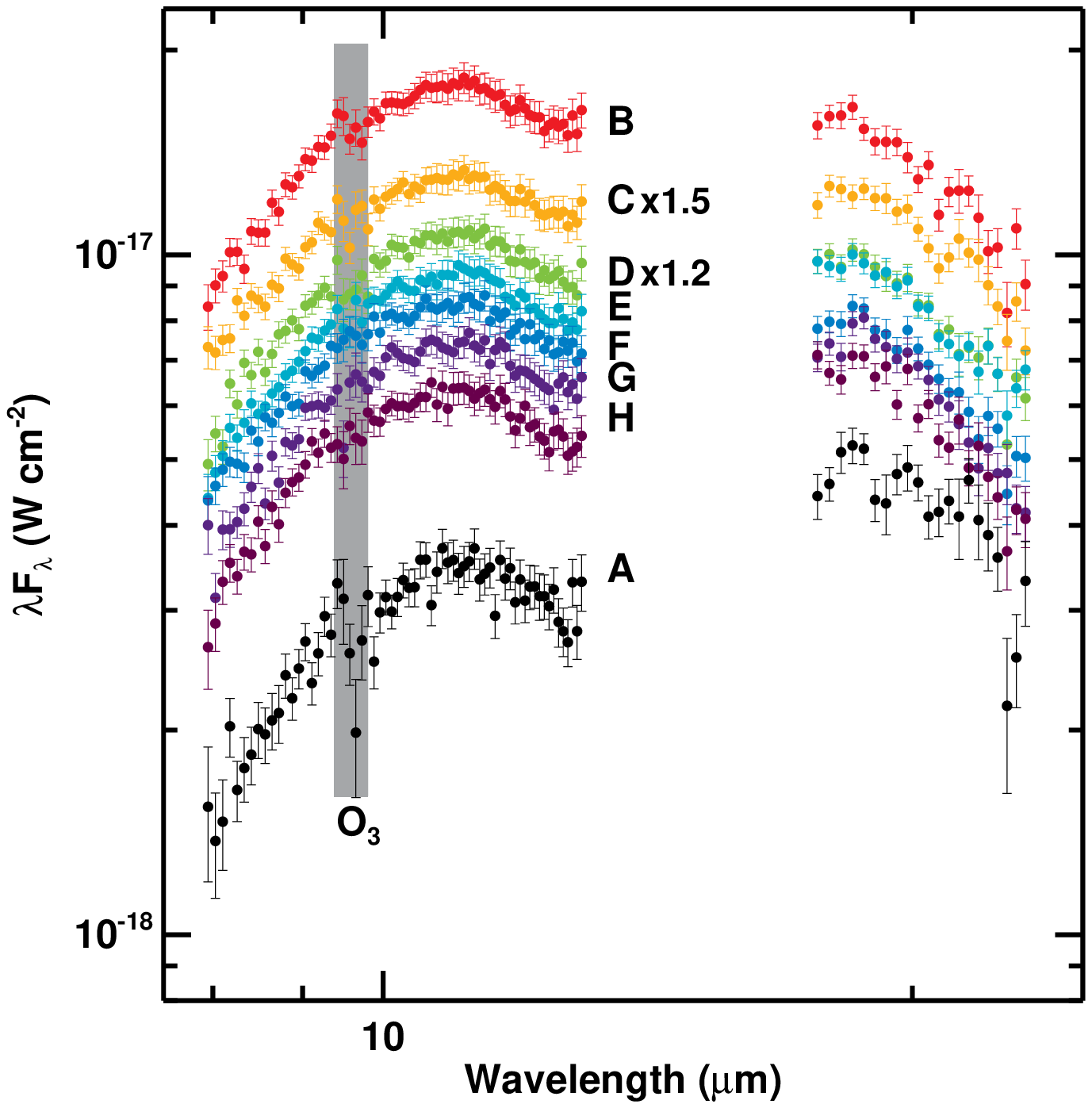}
\plottwo{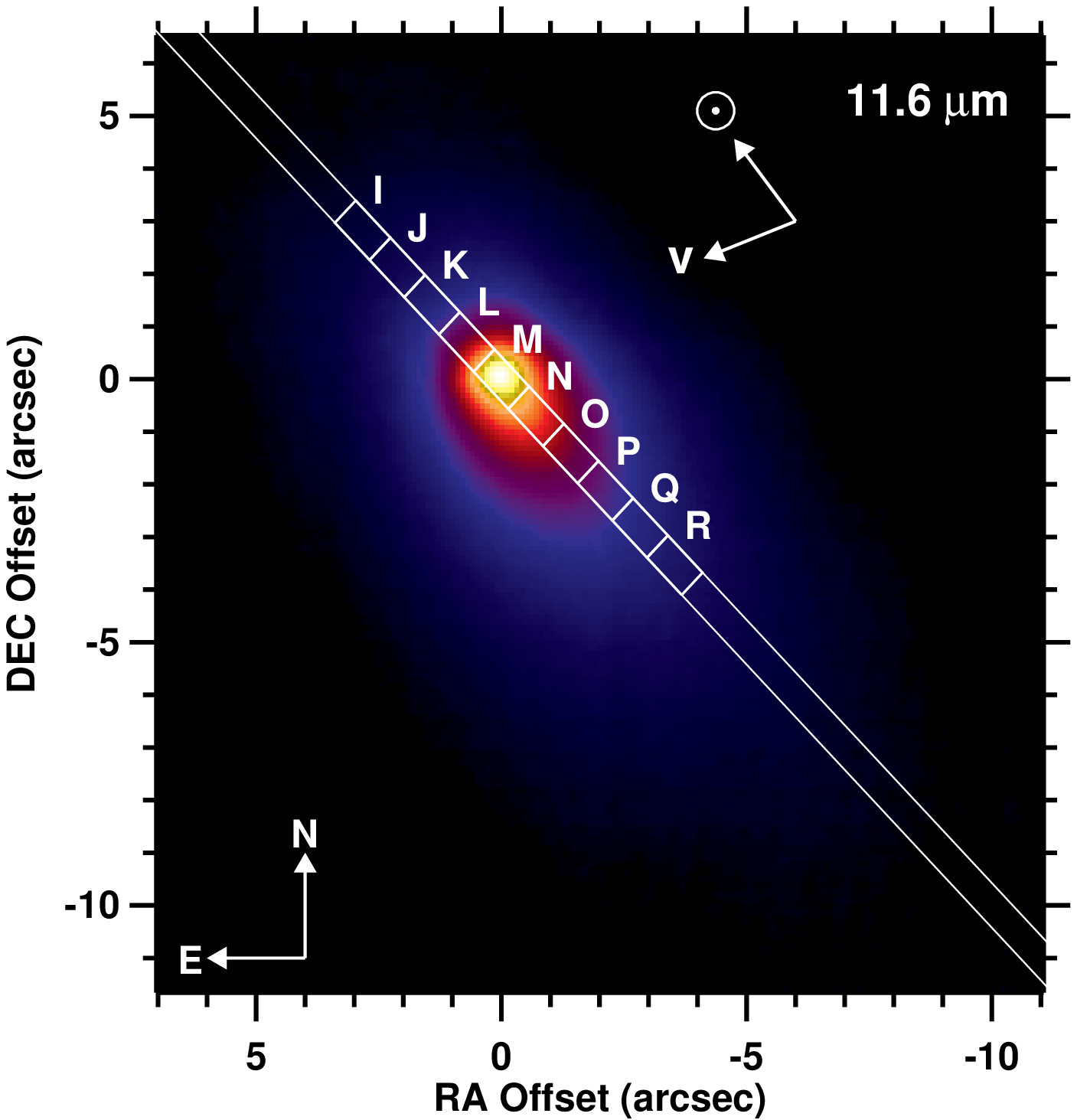}{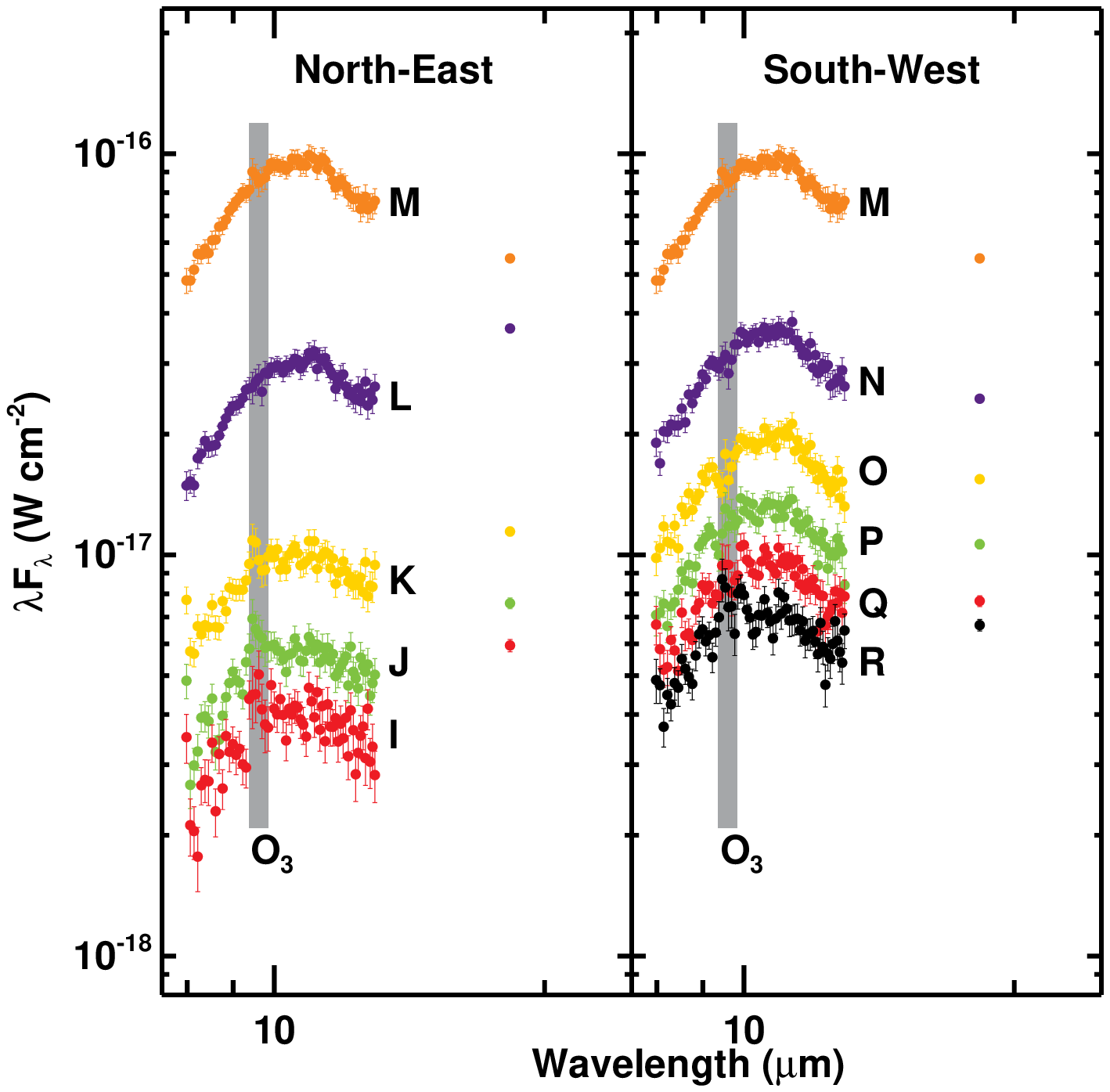}
\caption{Spectral extractions of comet 73P/Schwassmann-Wachmann~3 fragments 
SW3-[B] ({\it top}) and SW3-[C] ({\it bottom}).  The {\it left hand panels} 
are 11.6~\micron\ images of the fragments which show the location of the slit and the size and location of the spectral extraction apertures (labeled with
letters).  The {\it right hand panels} show the 10~\micron\ 
spectra for the fragments.
For SW3-[B], the spectral extractions for locations C and D have been 
multiplied by a factor of 1.5 and 1.2, respectively, for display purposes.
For SW3-[C], the extractions have been split to show extractions
North-East of the nucleus (which is at extraction point M, and is shown on
both sides for context) ({\it left}) and South-West of the nucleus ({\it right}).
Also shown in the {\it right hand panels} are the 20~\micron\ spectral extractions
for fragment SW3-[B] and the 18.1~\micron\ flux points for
fragment SW3-[C].
\label{fig:73pb_spec} }
\end{figure}

Figure~\ref{fig:73pb_spec} shows the location and orientation of the 
0.6\arcsec\ wide spectral slit during the observation of each fragment, the 
location and size of the extraction areas for the spectra, and the 
extracted spectra.  The slits were centered on the peak of the 
azimuthally-averaged coma surface brightness profile, essentially 
coincident with the nucleus (unresolved at our platescale), and rotated to 
align along the direction of the extended coma. 
Individual spectra were extracted in 1.0\arcsec\ segments in the spatial 
direction for mineralogical and grain analysis (\S\ref{sec:model}). 

The spectral extraction apertures centered on the nucleus contain both 
nuclear and coma fluxes. In order to study the coma dust properties, we need 
to properly assess the nuclear contribution to the flux.  We can 
deconvolve the relative contribution of the nucleus flux to the coma 
by analyzing the high spatial resolution images at 11.6~\micron.
We take the PSF of the standard star HD~127665 as the shape of the 
nuclear contribution to the images.  We assume a standard $1/\rho$ 
coma surface brightness decline. We derive the azimuthally-averaged 
cumulative photometry versus increasing $\rho$, i.e., from the peak flux 
pixel, for the image and for the sum of a scaled PSF and the $1/\rho$-coma. 
By $\chi^2$-fitting the shape of the 
azimuthially-averaged coma profile, we constrain the contribution of the PSF 
to the image.  By putting a synthetic beam of 0.6\arcsec\ $\times$ 
1.0\arcsec\ on this solution, the nucleus is found to contribute 
1.4\% $\pm$ 0.2\% of the measured flux in the centered slit at 11.6~\micron\ to a 
confidence level of 99\% for fragment SW3-[B].  Our calculated scaling
factor corresponds to an effective radius of the nucleus of 0.2~km $\pm$ 
0.05~km for fragment SW3-[B], equal to the lower limit of the nucleus radius
determined by \citet{howell07}.  
The nucleus contribution is subtracted from the central extracted spectrum 
of fragment SW3-[B], so we only need to model the remaining coma grains.
From the 11.6~\micron\ image of fragment SW3-[C], we determine that the flux
at the central region is produced solely from the surrounding coma, and that
there is no contribution to the flux from the embedded nucleus (at the 99\%
confidence level).  Therefore, we do not apply a correction to the spectrum for
fragment SW3-[C].

\section{MODEL OF THERMAL EMISSION FROM DUST GRAINS}
\label{sec:model}

The \citet{hark02} thermal emission dust code was used to model the 
observed dust grain emission in the coma of fragments SW3-[B] and SW3-[C]. 
The model assumes that an optically-thin collection of discrete (singular 
mineralogy) dust particles reside at the heliocentric ($r_{h}$) and 
geocentric ($\Delta$) distance of the comet at the epoch of observations 
(Table~\ref{tab:obssum}).  The 10~\micron{} mineralogy used in the model is 
derived from laboratory studies of interplanetary dust particles 
\citep[IDPs, ][]{woo00}, micrometeorites \citep{brad99}, the NASA
\textit{Stardust} mission \citep{brownlee06} and 
other remote sensing modeling efforts 
\citep[e.g.,][]{hanner94,hark02,woo04}. Our selection of basic dust grain 
components is consistent with the major mineral groups used by other 
modelers to generate synthetic SEDs arising from dust thermal emission in 
the 10~\micron\ region \citep{min05, sugita05, lisse06, werner06}.  
Amorphous silicates with chemical composition (stiochiometry) similar to 
olivine (Mg$_{y}$,Fe$_{(1-y)}$)$_{2}$SiO$_{4}$ and pyroxene 
(Mg$_{x}$,Fe$_{(1-x)}$)SiO$_{3}$ with $x = y = 0.5$ (i.e., $Mg/(Mg+Fe) = 0.5$)
reproduce the broad width of the 10~\micron{} feature.  Mg-rich orthopyroxene is 
detected through its 9.3 and 10.5~\micron{} features \citep{woo99,hark02}. Mg-rich 
crystalline olivine is uniquely identified through its distinct, relatively 
narrow 11.2~\micron{} silicate feature \citep{hanner94}. Mg-rich crystalline 
species are defined as grains with a stoichometry of $0.9 \le x \ltsimeq 1.0$ 
($Mg/(Mg+Fe) = 0.9$) \citep{woo08, koike03, chihara02}. 

Lastly, the presence of amorphous carbon grains in the coma is required in 
order to reproduce the underlying featureless emission (continuum) in the 
$8 - 13$~\micron{} wavelength region. Other species such as PAHs, hydrous 
silicates (clays) and iron sulfides \citep[see][]{werner06, gehrz07}, 
cannot be uniquely identified from our SW3 spectra. Therefore, these latter 
species are not included our model. 

The amorphous grains are assumed to be spherical (Mie Theory) and
range in size from $0.1 - 100$~\micron.  The indices of refraction
used in Mie Theory to calculate the optical efficiencies of each
amorphous silicate materials are from \citet{dors95} and the amorphous
carbon is from \citet{edoh83}.  The submicron sized crystalline
olivine grains are modeled using a continuous distribution of
ellipsoids (CDE) with one of the crystallographic axes elongated to a
ratio of 10:1:1 to match the wavelength of the emission features with
laboratory experiments \citep{fabi01}. The use of elongated grains is
supported by polarization measurements of the diffuse interstellar
medium which suggest the presence of elongated grains, and
condensation experiments which show the formation of elongated
forsterite (Mg-pure) crystals \citep{fabi01}. The radii of the
crystalline olivine grains ranges from $0.1 - 1.0$~\micron; larger
crystals do not well match the shape of the observed resonances in
comet Hale-Bopp \citep{min05}. The indices of refraction for the
three crystallographic axes of the crystalline olivine between 3 and
250~\micron\ are from \citet{jag98}.  The non-oriented crystalline
olivine indices ranging from $0.2 - 2$~\micron{} are from the Jena
Laboratory Group\footnotemark.
\footnotetext{http://www.astro.uni-jena.de/Laboratory/Database/databases.html}
The short wavelength indices are appended to each long wavelength axes
prior to the CDE calculations.  Note that the radius of the crystals
is an ``effective'' radius, or the radius a crystal would be if it
were spherical.

The temperature of each grain is calculated using a thermal
equilibration calculation.  In the coma, the grains reach radiative
equilibrium quickly \citep{woo79} as they enter into the coma.  The
temperature of the crystalline olivine grains are calculated using the
``hot crystal model'' presented in \citet{hark02}: the temperature of
the crystalline olivine grains are increased by 1.9 times over those
predicted from the optical constants in order to fit the
\textit{Infrared Space Observatory} (ISO) SEDs from comet
C/1995~O1~(Hale-Bopp) at 2.9~AU \citep{crovis97}.  The temperature
increase was required to match the relative emission peak strengths in
the 10 and 20~\micron\ regions, and may be caused by crystals being in
contact with warm material, or the crystals themselves contain some Fe
making them Mg-rich, but not Mg-pure.  We retain this ``hot crystal
model'' in all of our comet modeling to compare results between
comets, although the model awaits further constraining with additional
long wavelength data from the comets observed by \textit{Spitzer}\,
\citep{kelley06}.  An effect of using warm crystals in thermal models
is the calculated dust species masses derived from mineral
decomposition analysis are less than the mass fractions computed if
cooler temperature grain populations were invoked.  However, 
use of warm crystals does not affect the relative amount of crystalline
silicates we calculate in the fragment comae of SW3.

For the grain size distribution, we use the Hanner modified power law
\citep[hereafter HGSD;][]{hanner94}.  In the HGSD, the relative
number of each grain {\it size} is calculated by:
$n(a) = (1 - a_{\circ}/a)^M (a_{\circ}/a)^N$, where $a$ is the grain
radius, $a_{\circ} = 0.1$~\micron, the minimum grain radius, and $M$
and $N$ are independent parameters.  Since $N$ weights more of the
larger radii particles and can be considered as the size distribution
slope for large radii, it is best constrained at wavelengths
longer than 13~\micron.  The value for $N$ is set at three different
values in the model: 1)~$N = 3.4$ which weights more of the larger
particles in the size distribution; 2)~$N = 3.7$ is the canonical
value we determined from our modeling of comet C/1995~O1 (Hale-Bopp)
\citep{hark02}; and 3) $N = 4.2$ which weights more of the smaller
particles in the size distribution. From the values of 
$N$ and $M$, we also calculate
the peak of the HGSD: $a_{p} = (M + N)/N$.  The value $a_{p}$ is not a
freely varying parameter, but a calculated property of the size distribution.  
We varied $M$ such that $a_{p}$ changed by increments of 0.1~\micron.

The amorphous dust can be modeled as porous grains (the crystals always 
remain solid).  The fractal porosity of the amorphous grains is varied to 
create porous or ``fluffy'' particles \citep{lisse98}.  The amount of 
vacuum is: $f = 1 - (a/a_{\circ})^{(D - 3)}$ with $D = 3.0$ corresponding 
to solid grains and $D = 2.5$ corresponding to the maximum fractal 
porosity allowed while still satisfying the assumptions of Mie Theory 
\citep{baze90}. 

Finally, the flux (in W cm$^{-2}$ \micron$^{-1}$) from each mineral is 
calculated by integrating over the
grain size distribution between the minimum radius $a_{\circ} = 0.1$~\micron\ 
and the maximum grain radius $a_{max} = 100$~\micron,

\begin{equation}
F_{em}(\lambda,r_{h}) = \frac{N_p}{\Delta^{2}} \int_{a_{\circ}}^{a_{max}} 
n(a) \pi a^{2} Q_{em}(a,\lambda) B(\lambda,T_{d}(a)) da 
\label{eq:flux}
\end{equation}

\noindent where $N_{p}$ is the number of particles at the peak of the grain 
size distribution if the peak of $n(a)$ is normalized to unity, $\Delta$ is 
the geocentric distance of the dust grain, $n(a)~da$ is the grain size 
distribution, $Q_{em}(a,\lambda)$ is the emission efficiency for a grain of 
radius $a$(\micron) at wavelength $\lambda$(\micron), and $T_{d}$(K) is the 
heliocentric distance dependent ($r_{h}$-dependent) dust temperature for a 
single grain size, $a$, calculated by radiative equilibrium. The flux from 
each mineral species is coadded and fit to the Gemini spectra by varying 
$N_{p}$.  The best-fit is determined by calculating the lowest 
$\chi^{2}_{\nu}$ (reduced $\chi^{2}$) value at a confidence level of 
two-sigma.

\section{RESULTS \& DISCUSSION}
\label{sec:results}

\subsection{{\it Model Fits}}
\label{sec:fits}

\begin{figure}
\epsscale{0.65}
\plotone{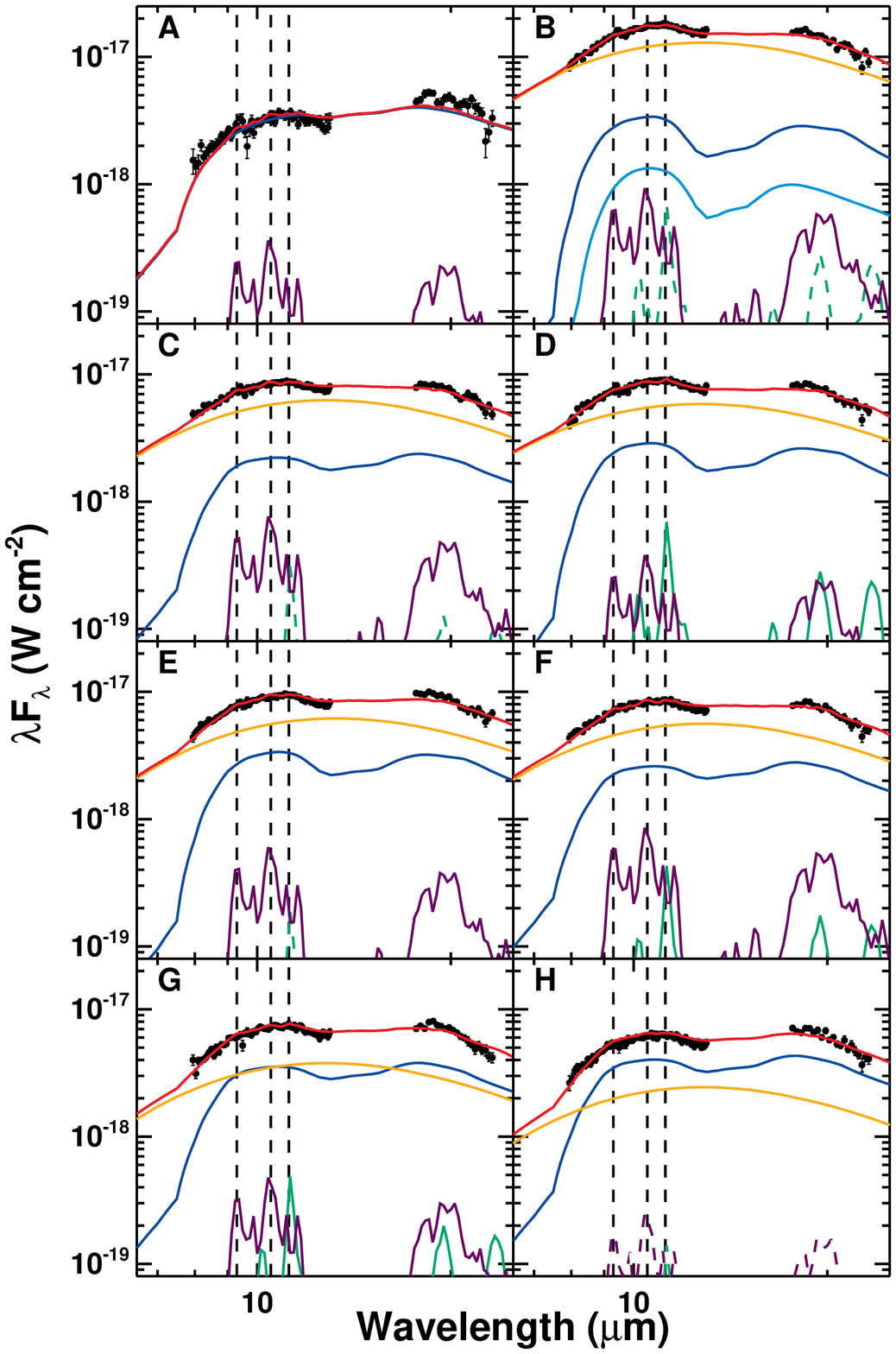}
\caption{Spectral decomposition of model fits to the coma thermal spectra 
of comet 73P/Schwassmann-Wachmann~3 fragment SW3-[B]. The location of the 
extraction beam is indicated in each panel.  Overlayed on the spectra ({\it 
black circles}) is the total SED fit ({\it red line}).  The minerals used 
to create the SED are: amorphous pyroxene ({\it blue line}), amorphous 
olivine ({\it cyan line}), amorphous carbon ({\it orange line}), 
crystalline olivine ({\it green line}), and orthopyroxene ({\it violet line}). 
A solid line for a particular mineral indicates that the mineral is constrained by 
the model, a dashed line indicates that the mineral is not constrained.  The
three vertical {\it black dashed lines} indicate the wavelength locations for
the resonance features of orthopyroxene at 9.3 and 10.5~\micron{} and for
crystalline olivine at 11.2~\micron. 
\label{fig:modelb} }
\end{figure}

\begin{figure}
\epsscale{0.5}
\plotone{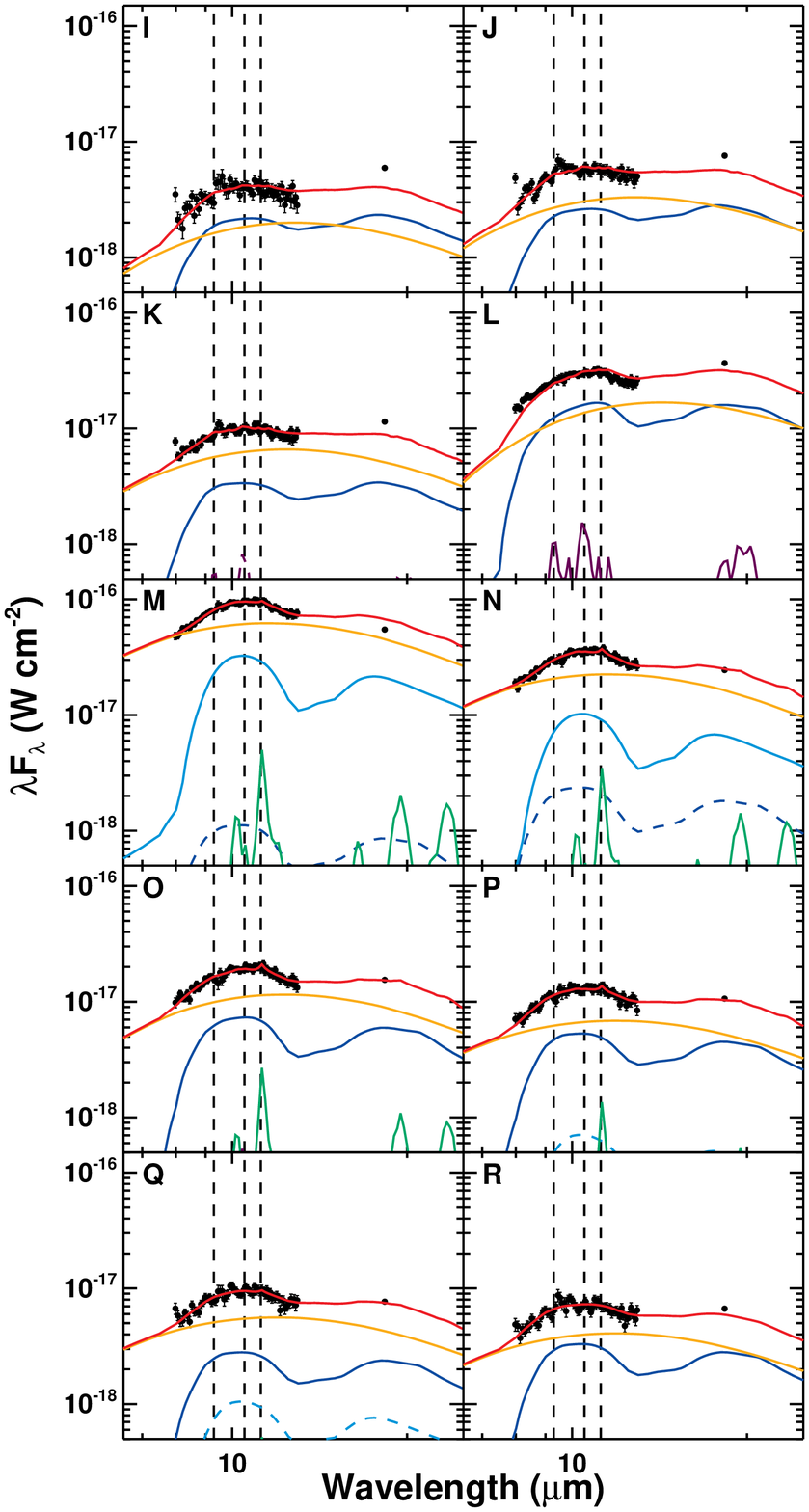}
\caption{Spectral decomposition of model fits to the coma thermal spectra 
of comet 73P/Schwassmann-Wachmann~3 fragment SW3-[C]. The location of the 
extraction beam is indicated in each panel.  Overlayed on the spectra ({\it 
black circles}) is the total SED fit ({\it red line}).  The minerals used 
to create the SED are: amorphous pyroxene ({\it blue line}), amorphous 
olivine ({\it cyan line}), amorphous carbon ({\it orange line}),
crystalline olivine ({\it green line}), and orthopyroxene ({\it violet line}).
A solid line for a particular mineral indicates that the mineral is constrained by 
the model, a dashed line indicates that the mineral is not constrained.  The
three vertical {\it black dashed lines} indicate the wavelength locations for
the resonance features of orthopyroxene at 9.3 and 10.5~\micron{} and for
crystalline olivine at 11.2~\micron. 
\label{fig:modelc} }
\end{figure}

The best-fit model parameters for SW3-[B] are listed in 
Table~\ref{tab:modparams_b} and the spectral decomposition is shown in 
Fig.~\ref{fig:modelb}.  Similarly, the best fit model parameters for SW3-[C] 
are listed in Table~\ref{tab:modparams_c} and the spectral decomposition 
is shown in Fig.~\ref{fig:modelc}. 

The grain properties of both SW3 fragments are similar.  The
grains range from being solid to moderately fractally porous ($D = 3.0
- 2.727$).  Fractally porous coma grains in SW3 fragments
are also required to explain the high fractional polarization
measured in the optical $I$- and near-IR $H$-band at high phase angles
and throughout the comae of SW3-[B] and SW3-[C] by \citet{tjjones08}.  The peak 
of the HGSD for fragment SW3-[B] in the anti-sunward direction
is slightly larger ($\sim 0.5$~\micron) than that of fragment SW3-[C]
($\sim 0.3$~\micron).  However, the grain
population in SW3 overall is not dominated by extremely small particles as was
evident in the comae C/1995~O1 (Hale-Bopp) or 9P post-Deep Impact
\citep{woo99,hark05,hark07}.

In both fragments, most of the thermal emission is dominated by
amorphous (glassy) carbon and silicates.  The silicate emission
feature in both fragments is dominated by amorphous pyroxene grains,
although there is some evidence for a population of amorphous olivine
grains associated with the coma toward the nuclei of both
fragments. Orthopyroxene is seen in the extended coma in both the
sunward and anti-sunward direction of fragment SW3-[B], but only in the sunward
direction, close to the nucleus (1\arcsec) in fragment SW3-[C].  A modest 
population of crystalline
olivine grains is present in both fragments. The silicate crystalline
mass fraction for the submicron to micron-size portion of the grain
size distribution \citep{hark02,moreno03}, defined as
$f_{cryst} \equiv$ (crystalline)/(crystalline~+~amorphous), 
for SW3-[B] ranges from $0.335^{+0.089}_{-0.112}$ at the nucleus, 
declining to $0.061^{+0.068}_{-0.061}$ at a cometocentric distance of 
6.0\arcsec{} out into the coma in the anti-sunward direction. 
The $f_{cryst}$ derived for SW3-[C] is $0.257^{+0.039}_{-0.043}$ 
towards the nucleus, declining to $0.0^{+0.073}_{-0.0}$ at a 
cometocentric distance of 5\arcsec{} into the coma
in the anti-sunward direction.   Therefore, within the 1-sigma
confidence level, both fragments have a similar crystalline silicate
fraction.  For many EC
comets and in some NICs, the fraction of cometary grains that are
Mg-rich is high ($f_{cryst} \simeq 0.08 - 0.3$). In 9P, $f_{cryst}
\simeq 0.3$ was found in the post-impact coma \citep{hark07, sugita05}. 
Overall, lack of significant mineralogical differences
between SW3 fragments may indicate that the original nucleus of the
comet was generally homogeneous in composition as opposed to that of
1P/Halley or 9P \citep{woo08}.

The spectra of fragment SW3-[C] in the sunward direction
are dominated by emission from amorphous pyroxene and carbon, with
little to no emission from crystals (Table~\ref{tab:modparams_c}).  The peak of the 
grain size distribution is also slightly larger (0.5 --1.3~\micron) compared 
with the derived values in the anti-sunward direction (0.3 -- 0.4~\micron), in addition to
being more solid than the latter.  We attribute these differences in the grain properties 
to the dynamics of the grains as they are released from the nucleus of
fragment SW3-[C].  Smaller and/or more porous grains which can exhibit
crystalline emission are more easily affected by the radiation pressure or
are better coupled to and entrained in the sublimating gas (e.g., \citep{hark07,orosei1995}).
Likely these grains are more easily
swept out into the anti-sunward direction.  However, a full dynamical analysis 
of the trajectory of the grains is outside the scope of this work.

The relative abundance of amorphous carbon in the anti-sunward coma of
SW3-[B] ranges between $\simeq 45 - 23$~wt\%, while for SW3-[C] the value
ranges between $\simeq 60 - 42$~wt\%. The wt\% of amorphous carbon 
in both fragments varies a little with distance from the nucleus 
($\simeq 20$\%) and is comparable to mean values deduced for other
comets, especially ECs (10 to 50~wt\%) comae \citep{woo08}. For
example \citet{hark02, hark04} find amorphous carbon abundance
(relative to other refractory materials) of 21~wt\% for comet
C/1995~O1 (Hale-Bopp), \citet{woo04} cite 15~wt\% for
C/2001~Q4~(NEAT), while \citet{hark07} deduce a 28~wt\% relative
abundance for the pre-impact coma of 9P. Each of the later values
(including those for SW3) were derived using the
same discrete composition thermal model approach for SED
decomposition, enabling a consistent comparison of relative abundances
that are not model dependent \citep{woo08}. 

The average silicate-to-carbon ratio in the extended anti-sunward 
coma for fragment SW3-[B] ranges between $1.156 - 3.43$, while for 
fragment SW3-[C] the value ranges between $0.671 - 1.882$. Overall, these
ratios are comparable to values derived for other EC comets 
\citep{kelley06, hark07, lisse06, reach10}. 
Fragment SW3-[B] went through an outburst event just prior 
to our observations and this event could be responsible for the 
higher silicate-to-carbon ratio in fragment SW3-[B] compared to that 
of fragment SW3-[C], as well as the slight differences in the
peak of the grain size distribution between the two fragments.  The date of the 
outburst from fragment SW3-[B] has been estimated to have been between 
2006 April 16 to 26  \citep{vincent2010}.  Likely,  material from the 
outburst is responsible for the region of enhanced flux in fragment SW3-[B] seen
3\arcsec\ from the nucleus region in our images.  The activity from fragment SW3-[C]
is primarily from two active jets on the nucleus \citep{vincent2010} that
could be preferentially ejecting small grains from the nucleus subsurface.
On the other hand, fragment SW3-[B] was exhibiting fragmentation
that released all subsurface grains in a single event.  Therefore, in bulk, the
grains released during a fragmentation event are slightly larger and
more silicate rich than those entrained in the active jet outflows from the
nucleus surface.

The radial difference in the silicate-to-carbon ratio in each fragment could arise 
from fragmentation of composite silicate rich grain aggregates 
entrained in the out flowing gas.  \citet{tjjones08} observed 
fragments SW3-[B] and SW3-[C] in imaging polarimetry mode in the near-IR,
finding that the surface brightnesses of both fragments are consistent 
with significant grain fragmentation. Their model requires a factor 
of 10 change in the mean grain size within 2--200~km from the 
nucleus (corresponding to 2\arcsec{} in our Michelle observations).  
In addition, the scattered light polarization profile slightly increased 
over the same distances. The latter observations suggest that the grains were
fragmenting, although the polarization could have also increased
from the release of volatiles such as water ice and organics, which
may change the light scattering properties of the grains
\citep{tjjones08}. Despite the evidence for fragmentation in the
scattered light, the grain size distribution derived from our
thermal models do not appreciably vary, nor does $a_{p}$.
Our best-fit grain size distributions have $N=3.4$.
This shallow power-law slope suggests that the scattered light could
be dominated by the largest grains which are too cool to be seen in the
10~\micron{} region.  Because smaller grains have
higher radiative equilibrium temperatures, the mid-IR emission at
10~\micron{} may be dominated by $\sim0.1-1$~\micron{} grains
\citep{reach07}.  Therefore the two datasets may not be so
incongruent if the largest grains, perhaps 10s or 100s of \micron{}
in size, are fragmenting, but do not significantly affect the 8--13
SED because of their cool radiative equilibrium temperatures.

Alternatively, the increase in silicate-to-carbon ratio with
distance from the nucleus could be due to grain sorting in the coma.
In this scenario, the more transparent silicate species would
dynamically separate from the dark amorphous carbon grains.
Moderately porous ($D=2.857$) amorphous carbon grains have a 2--3
times larger response to radiation pressure than do the same-sized
silicate grains \citep{kelley06phd}.  Grain sorting by size and
mineral composition was suggested from the time-of-flight analysis of
the 9P post-impact coma contaminated by material from the ejecta plume
\citep{hark07}.

\subsection{{\it Comparison to Other Comets}}
\label{sec:disc_sc}

\begin{figure}
\epsscale{0.65}
\plotone{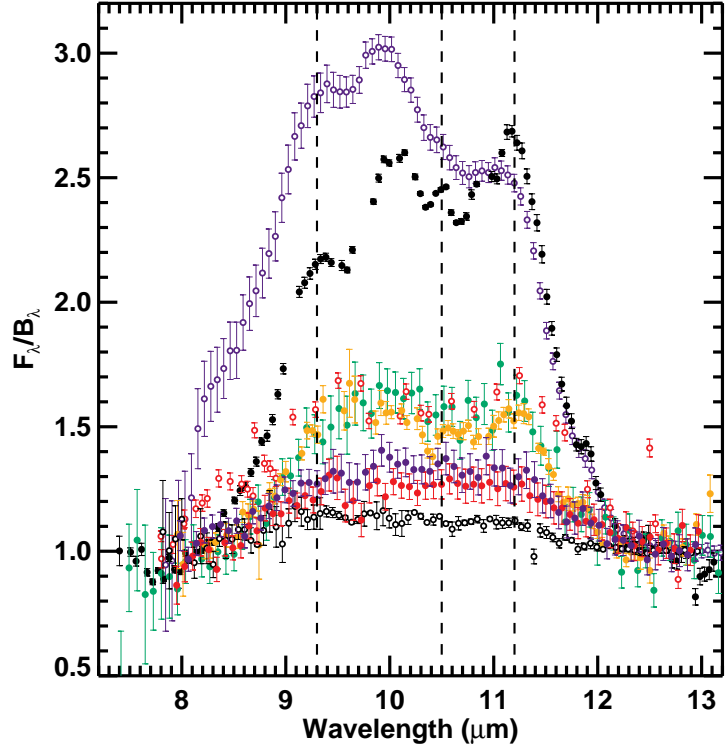}
\caption{Blackbody normalized comet spectra extracted from slit positions 
coincident with each nucleus.  The normalization is derived from fits to 
continuum spectral points $< 8.2$~\micron{} and $> 12.5$~\micron{} lying 
outside the silicate band.  The blackbody normalized spectrum of comet 
73P/Schwassmann-Wachmann~3 fragments SW3-[B] ({\it red}; 
B$_{\lambda}$(T) $= 289$~K, r$_h = 1.11$, $\Delta = 0.15$) and SW3-[C] 
({\it blue}; B$_{\lambda}$(T) $= 298$~K, r$_h = 1.09$, $\Delta = 0.13$) 
10~\micron\ spectra are compared to other ecliptic (Jupiter family) and nearly
isotropic (Oort cloud) comets observed with similar ground-based
remote sensing techniques (except for C/2004~B1 (LINEAR) and 17P/Holmes
which were observed with {\it Spitzer} and 1P/Halley which was observed with
the Kupier Airborne Observatory): C/2004 B1 (Linear) ({\it open black circles}; 
B$_{\lambda}$(T) $= 215$~K, r$_h = 2.06$, $\Delta = 1.60$), C/2001 Q4 (NEAT) 
({\it orange}; B$_{\lambda}$(T) $= 310$~K, r$_h = 0.97$, $\Delta = 0.35$), 
9P/Tempel-1 post-Deep Impact ({\it green}; B$_{\lambda}$(T) $= 288$~K, r$_h = 1.51$, 
$\Delta = 0.89$), 1P/Halley ({\it open red circles}; B$_{\lambda}$(T) $= 300$~K, 
r$_h = 1.32$, $\Delta = 0.77$), C/1995 O1 (Hale-Bopp) ({\it black}; 
B$_{\lambda}$(T) $= 288$~K, r$_h = 0.93$, $\Delta = 1.46$), and
17P/Holmes ({\it open blue circles}; B$_{\lambda}$(T) $= 206$~K, 
r$_h = 2.51$, $\Delta = 1.87$)  The {\it dashed lines} indicate the wavelengths
of crystalline silicate emission features at 9.3 and 10.5~\micron{} (orthopyroxene),
and 11.2~\micron{} (crystalline olivine).
\label{fig:comp} }
\end{figure}

It is instructive to compare the spectral signature of coma dust in
the environs of the SW3 fragments to other comets observed at mid-IR
wavelengths to place the conclusion based on thermal model analysis
and mineral decomposition analysis in context.  Fig.~\ref{fig:comp} is
a composite plot of the spectra of fragments SW3-[B],
SW3-[C], comets C/1995~O1 (Hale-Bopp), 9P post-Deep Impact, and
C/2001~Q4 (NEAT) observed with ground-based mid-IR spectrographs, 
C/2004 B1 (LINEAR) and 17P/ Holmes observed with {\it Spitzer} IRS spectrograph;
and 1P/Halley observed by the Kupier Airborne Observatory. 
Comets C/1995~O1 (Hale-Bopp) and 17P/Holmes (17P) clearly have the strongest silicate 
features (defined as the excess over the blackbody continuum).  However, we note
that 17P was observed by {\it Spitzer} a little more than a couple of weeks after 
an extremely violent outburst event.  This material was still within the
{\it Spitzer} IRS spectral beam, and is therefore the reason for the 
relatively large silicate feature.  The grains in
the coma of C/1995~O1 (Hale-Bopp) were fractally very porous ($D = 2.5$) with the
HGSD peaking at $a_{p}=0.2$~\micron{} \citep{hark02}.  The coma dust
in C/1995~O1 (Hale-Bopp) is also mineralogically diverse and includes amorphous
and crystalline forms of both olivine and pyroxene \citep{hark02}.
Similarly, the dust ejected from 9P post-Deep Impact
was also mineralogically diverse and small ($a_p = 0.2$~\micron) in
size, although the grains were moderately fractally porous ($D =
2.727$) \citep{hark05}.  Furthermore, comet C/2001~Q4 (NEAT) also
displayed a mineralogically diverse grain population, however, the grains
were solid ($D = 3.0$) and and slightly larger than those in Hale-Bopp and
9P ($a_p = 0.3$~\micron) \citep{woo04}.

Comet 17P/Holmes also was observed to have distinct Mg-rich crystalline 
olivine features at 11.2~\micron{} and 11.9~\micron{} in apertures 
centered on the nucleus and in the diffuse coma, as well as toward the 
isolated dust cloud arising from the violent outburst event \citep{reach10, 
watanabe10}.  In 17P, dust producing these 10~\micron{} features may 
arise from fluffy aggregate complexes containing small Mg-rich 
crystalline silicates \citep[e.g.,][]{kimura08}, or small silicates ($a 
\ltsimeq 0.1$~\micron{}) admixed larger amorphous dust. Sharp 
crystalline silicate features from small grains superposed on a thermal 
continuum arising from larger amorphous carbonaceous and silicate 
species in the 10~\micron{} SED were evident in remote 
sensing observations of comet 1P/Halley by the Kupier Airborne Observatory
\citep{bregman87}, while the 
\textit{in situ} flyby measurement through the coma \citep{mcd87} 
indicated that the grain sized distribution was dominated by large 
$\simeq 1$~\micron{} sized particles.  Comet C/2004 B1 (LINEAR) has a
relatively weak and featureless silicate feature indicative of emission from
a large ($> 1$\micron) grain population.

While there are similarities in dust grain characteristics between
C/1995~O1 (Hale-Bopp), 9P, and 17P possibly a signature of a common
formation zone in the solar nebula of the parent body aggregates for
these comets, the composition of SW3 seems distinctly different. As
SW3 fragmented into may pieces that subsequently
sublimated leading to outbursts of gas entrained with dust
particles from within the bulk interior of the original nucleus
(natural excavation at depth akin to the subsurface penetration of
Deep Impact projectile), surface weather affects cannot be invoked to
explain dissimilarities in grain properties. ECs with frequent
perihelion passages and low semi-major orbital axis are thought to
have coma grain populations dominated by larger, compact dust
particles arising from highly processed surface layers constituting
the nucleus crust  \citep{prialnik91} as opposed
to the more pristine grains released in active Oort cloud
comets. Larger grains are necessary to explain the aperture-dependent
polarization properties and low total polarization ($\ltsimeq$ few
percent) of some ECs \citep{tjjones08,koloko07}. However, SW3 has an
unusually high polarization for an EC
in both the optical and near-IR at a range of phase angles, which lead
\citet{tjjones08} to suggest that
observational feature is best explained by the release of unprocessed,
porous aggregates with small effective radii from the interior of
disintegrating comet nuclei.

Although fragment SW3-[B] underwent an outburst event revealing pristine
grains from the nucleus interior, the grain properties and mineralogy 
of its dust does not resemble that of either 9P or of Hale-Bopp.  This 
dissimilarity could be attributed to 1)~the smaller, more porous dust 
grains traveling out of the coma of fragment SW3-[B] prior to our observations; 
or 2)~the dust properties of SW3 \textit{fundamentally differ} from that of 
9P (and Hale-Bopp) implying a different formation process or evolutionary
history. However, the former conjecture is ruled out by polarimetric
observations and dust fragmentation studies of the SW3-[B] coma
\citep{tjjones08} obtained at epochs contemporaneous with our mid-IR
spectrophotometry.

\section{CONCLUSIONS}
\label{sec:concl}

Mid-IR images and spectroscopic observations of the ecliptic (Jupiter family)
comet 73P/Schwassman-Wachmann~3, fragments SW3-[B] and SW3-[C], were analyzed
using thermal emission models.  Based on
our modeling, the comae dust, and by inference the bulk internal
properties, of the individual fragments are similar with 
the observed mid-IR flux being dominated by emission from 
amorphous pyroxene and amorphous carbon, with evidence for emission from
amorphous olivine towards the nucleus in both fragments.  Emission from 
both fragments shows evidence for the existence of some crystalline olivine 
in the extended coma in the anti-sunward direction.  The grain properties in 
both fragments are also similar: solid to moderately porous grains 
($D = 3.0 - 2.727$) with a large grain slope of $N = 3.4$ and
the peak of the HGSD ranging between 0.3 and $\sim 1.0$~\micron.
Finally, the two fragments exhibit a similar, but slightly different
silicate-to-carbon ratio (1.341 for SW3-[B] and
0.671 for SW3-[C]); a ratio equivalent to those found for other ECs.
The slight differences in grain size distribution and silicate-to-carbon 
ratio between the two fragments likely arises because SW3-[B] was actively fragmenting
throughout its passage while the activity in SW3-[C] was primarily driven by
jets.
The breakup event populated the coma of fragment SW3-[B] with subsurface material.
Conversely, for fragment SW3-[C], the dust population of its coma was driven
by two active jets.

There are other differences between the two fragments, including
the detection in fragment SW3-[B] 
of orthopyroxene in both the sunward and anti-sunward directions.
In fragment SW3-[C], orthopyroxene is primarily detected in only the sunward
direction. The crystalline silicate fraction is higher in SW3-[B] than in SW3-[C]
based on constraints derived for the crystalline silicate fraction from
thermal modelling of the SEDs.  Although $f_{cryst}$ for both fragments 
is similar to the range of $f_{cryst}$ inferred for other ECs.

Comparison of coma dust properties of 73P to those of derived for ecliptic 
comet 9P/Tempel~1 just after the Deep Impact event, and to the archetypal, 
``pristine'' nearly-isotropic (Oort cloud) comet C/1995~O1 (Hale-Bopp) 
suggests that materials released from fragment SW3-[B] 
which may expose grain material buried at depth below the surface of the 
original larger nucleus is devoid of very small ($a \leq 0.5$~\micron) 
dust grains. The apparent absence, or diminished population of
such small grain species may either be a result of dynamical sorting
whereby this population has passed out of the coma before our observations, or 
more likely, the initial parent body makeup of 73P is different than 
that of 9P/Tempel~1 or even C/1995~O1 (Hale-Bopp).

\acknowledgements Data discussed in this manuscript are based on
observations obtained at the Gemini Observatory, which is operated by
the Association of Universities for Research in Astronomy, Inc., under
a cooperative agreement with the NSF on behalf of the Gemini
partnership: the National Science Foundation (United States), the
Science and Technology Facilities Council (United Kingdom), the
National Research Council (Canada), CONICYT (Chile), the Australian
Research Council (Australia), Minist{\'e}rio da Ci{\^e}ncia e
Tecnologia (Brazil) and Ministerio de Ciencia, Tecnolog{\'i}a e
Innovaci{\'o}n Productiva (Argentina). DEH and CEW acknowledge support
for this work from the National Science Foundation grant
AST-0706980. DEH, DHW and CEW also acknowledge partial support for
this work from NASA Planetary Astronomy Grant RTOP 344-32-21-04.  MSK
acknowleges support from NASA Planetary Astronomy Grant NNX09AF10G.
The authors also would like to thank the Gemini Observatory staff for
their support in conducting these observations as well as an anonymous
referee whose comments improved the manuscript.

\newpage


\clearpage

\begin{deluxetable}{lcccccccccc}
\rotate
\tablewidth{0pc}
\tablecaption{MICHELLE OBSERVATIONAL SUMMARY \label{tab:obssum}}
\tablehead{
\colhead{UT} & \colhead{Start Time \tablenotemark{a}} & &
\colhead{Airmass} & \colhead{Int.\ Time\tablenotemark{b}} & \colhead{Flux} & \colhead{Airmass of} & 
\colhead{$r_h$} & \colhead{$\Delta$} & \colhead{Data} & \colhead{Flux \tablenotemark{f}}\\
\colhead{Date} & \colhead{(hr:min)} & Frag. &
\colhead{of Comet} & \colhead{(sec)} & \colhead{Standard} & \colhead{Standard} & 
\colhead{(AU)} & \colhead{(AU)} & \colhead{Obtained \tablenotemark{c}} & \colhead{(Jy)}
}
\startdata
29 Apr 2006 & 12:21 & [B] & 1.039 & 54.4 & HD127665 & 1.065 & 1.110 & 0.154 & 18.1~\micron\ image & $13.60 \pm 0.33$\\
29 Apr 2006 & 12:23 & [B] & 1.043 & 51.8 & HD127665 & 1.072 & 1.110 & 0.154 & 11.6~\micron\ image & $6.54 \pm 0.07$ \\
29 Apr 2006 & 12:41 & [B] & 1.104 & 806.4 & HD127665 & 1.096 & 1.110 & 0.154 & Low N\tablenotemark{d} & $\cdots$ \\
29 Apr 2006 & 13:55 & [B] & 1.199 & 756.0 & HD156283 & 1.151 & 1.110 & 0.154 & Low Q\tablenotemark{d} & $\cdots$ \\

30 Apr 2006 & 11:36 & [C] & 1.043 & 54.4 & HD156283 & 1.151 & 1.085 & 0.129 & 18.1~\micron\ image & $41.64 \pm 1.00$ \\
30 Apr 2006 & 11:42 & [C] & 1.039 & 51.8 & HD156283 & 1.048 & 1.085 & 0.129 & 11.6~\micron\ image & $30.10 \pm 0.30$ \\
30 Apr 2006 & 12:57 & [C] & 1.034 & 302.4 & HD156283 & 1.056 & 1.085 & 0.129 & Low N\tablenotemark{e} & $\cdots$ 

\enddata
\tablenotetext{a}{ \ Start time of integration.}
\tablenotetext{b}{ \ Integration time on-source.}
\tablenotetext{c}{ \ Low N $=$ Low-Res 10~\micron\ spectrum; 
Low Q $=$ Low-Res 20~\micron\ spectrum.}
\tablenotetext{d}{ \ Slit rotated to position angle $= 25^{\circ}$}
\tablenotetext{e}{ \ Slit rotated to position angle $= 45^{\circ}$}
\tablenotetext{f}{ \ Flux in a 7.44\arcsec\ diameter aperture centered on the peak isophote of the surface brightness distribution.}

\end{deluxetable}

\begin{deluxetable}{cccccccccccc}
\rotate
\tablewidth{0pc}
\tablecaption{BEST-FIT THERMAL EMISSION MODEL PARAMETERS FOR FRAGMENT B \label{tab:modparams_b}}
\tablehead{
 & & & & & & \multicolumn{5}{c}{$N_p (\times 10^{16}$)} & \\
\colhead{Extraction\tablenotemark{a}} & \colhead{Offset \tablenotemark{b}} & \colhead{N} & \colhead{M} &
\colhead{$a_p$ \tablenotemark{c}} & \colhead{D} & \colhead{Amorphous} &
\colhead{Amorphous} & \colhead{Amorphous} & \colhead{Crystalline} & \colhead{Ortho-} &
\colhead{$\chi^{2}_{\nu}$} \\
\colhead{Aperture} & \colhead{(arcsec)} & & & & & \colhead{Pyroxene} & \colhead{Olivine} &
\colhead{Carbon} & \colhead{Olivine} & \colhead{Pyroxene} &
}
\startdata

A & $+1$ & 3.4 & 27.20 & 0.9 & 3.000 & ${\bf 0.643^{+0.009}_{-0.011}}$ & $0.000^{+0.003}_{-0.000}$ & $0.000^{+0.007}_{-0.000}$ & $0.000^{+0.047}_{-0.000}$ & ${\bf 0.854^{+0.233 }_{-0.233}}$ & 2.71 \\
B & $0$  & 3.4 & 13.60 & 0.5 & 2.727 & ${\bf 3.310^{+0.989}_{-0.989}}$ & ${\bf 0.782^{+0.461}_{-0.463}}$ & ${\bf9.703^{+0.327}_{-0.330}}$ & $0.686^{+0.701}_{-0.686}$ & ${\bf 2.648^{+1.338 }_{-1.311}}$ & 0.75 \\
C & $-1$ & 3.4 & 13.60 & 0.5 & 3.000 & ${\bf 2.671^{+0.312}_{-0.312}}$ & $0.000^{+0.036}_{-0.000}$ & ${\bf 5.296^{+0.213}_{-0.213}}$ & $0.268^{+0.310}_{-0.268}$ & ${\bf 1.701^{+ 0.554}_{-0.554}}$ & 1.14 \\
D & $-2$ & 3.4 & 10.20 & 0.4 & 2.857 & ${\bf 5.933^{+0.626}_{-0.639}}$ & $0.000^{+0.196}_{-0.000}$ & ${\bf 9.800^{+0.370}_{-0.369}}$ & ${\bf 0.980^{+0.439}_{-0.439}}$ & ${\bf 1.540^{+0.859 }_{-0.850}}$ & 1.04 \\
E & $-3$ & 3.4 & 13.60 & 0.5 & 2.857 & ${\bf 3.621^{+0.361}_{-0.367}}$ & $0.000^{+0.025}_{-0.000}$ & ${\bf 4.913^{+0.223}_{-0.223}}$ & $0.170^{+0.350}_{-0.170}$ & ${\bf 1.481^{+0.651 }_{-0.650}}$ & 1.55 \\
F & $-4$ & 3.4 & 13.60 & 0.5 & 3.000 & ${\bf 2.827^{+0.310}_{-0.309}}$ & $0.000^{+0.070}_{-0.000}$ & ${\bf 4.837^{+0.213}_{-0.214}}$ & ${\bf 0.393^{+0.306}_{-0.311}}$ & ${\bf 2.159^{+0.554 }_{-0.555}}$ & 1.00 \\
G & $-5$ & 3.4 & 13.60 & 0.5 & 3.000 & ${\bf 3.949^{+0.302}_{-0.302}}$ & $0.000^{+0.038}_{-0.000}$ & ${\bf 3.101^{+0.207}_{-0.208}}$ & ${\bf 0.409^{+0.285}_{-0.285}}$ & ${\bf 1.145^{+0.521 }_{-0.520}}$ & 1.38 \\
H & $-6$ & 3.4 & 13.60 & 0.5 & 3.000 & ${\bf 4.494^{+0.269}_{-0.277}}$ & $0.000^{+0.022}_{-0.000}$ & ${\bf 1.877^{+0.191}_{-0.191}}$ & $0.000^{+0.245}_{-0.000}$ & $0.384^{+0.474 }_{-0.384}$ & 1.46

\enddata

\tablenotetext{a}{ \ Lable of spectral extraction aperture (see 
Fig.~\ref{fig:73pb_spec})}
\tablenotetext{b}{ \  Distance that the center of the $0.^{\prime\prime}6 \times 1.0^{\prime\prime}$ extraction box is offset from the peak brightness.  A positive value indicates an offset in the N-E direction, and a negative value indicates an offset in the S-W direction (see Fig.~\ref{fig:73pb_spec}).  Both offsets are along the slit orientation axis of PA $= -25$ (see text).}
\tablenotetext{c}{ \ Derived parameter}
\tablenotetext{d}{ \ Bold values indicate values constrained to a confidence level of 1-sigma.}

\end{deluxetable}

\begin{deluxetable}{cccccccccccc}
\rotate
\tablewidth{0pc}
\tablecaption{BEST-FIT THERMAL EMISSION MODEL PARAMETERS FOR FRAGMENT C \label{tab:modparams_c}}
\tablehead{
 & & & & & & \multicolumn{5}{c}{$N_p (\times 10^{16}$)} & \\
\colhead{Extraction\tablenotemark{a}} & \colhead{Offset \tablenotemark{b}} & \colhead{N} & \colhead{M} &
\colhead{$a_p$ \tablenotemark{c}} & \colhead{D} & \colhead{Amorphous} &
\colhead{Amorphous} & \colhead{Amorphous} & \colhead{Crystalline} & \colhead{Ortho-} &
\colhead{$\chi^{2}_{\nu}$} \\
\colhead{Aperture} & \colhead{(arcsec)} & & & & & \colhead{Pyroxene} & \colhead{Olivine} &
\colhead{Carbon} & \colhead{Olivine} & \colhead{Pyroxene} &
}
\startdata

I & $+4$ & 3.4 & 13.60 & 0.5 & 3.000 & ${\bf 2.619^{+0.239}_{-0.241}}$ & $ 0.000^{+0.025}_{-0.000}$ & ${\bf   0.480^{+0.190}_{-0.194}}$ & $ 0.000^{+0.044}_{-0.000}$ & $0.000^{+0.225}_{-0.000}$ & 2.79 \\
J & $+3$ & 3.4 & 13.60 & 0.5 & 3.000 & ${\bf 3.000^{+0.285}_{-0.361}}$ & $ 0.000^{+0.029}_{-0.000}$ & ${\bf   1.287^{+0.251}_{-0.221}}$ & $ 0.000^{+0.042}_{-0.000}$ & $0.070^{+0.541}_{-0.070}$ & 2.58 \\
K & $+2$ & 3.4 & 10.20 & 0.4 & 3.000 & ${\bf 7.113^{+0.870}_{-0.887}}$ & $ 0.000^{+0.089}_{-0.000}$ & ${\bf   6.789^{+0.571}_{-0.567}}$ & $ 0.000^{+0.159}_{-0.000}$ & $0.903^{+1.081}_{-0.903}$ & 2.44 \\ 
L & $+1$ & 3.4 & 40.80 & 1.3 & 2.727 & ${\bf 0.692^{+0.078}_{-0.078}}$ & $ 0.000^{+0.002}_{-0.000}$ & ${\bf   0.483^{+0.049}_{-0.048}}$ & $ 0.000^{+0.114}_{-0.000}$ & ${\bf 4.747^{+2.902}_{-2.888}}$ & 4.00 \\
M & $ 0$ & 3.4 &  6.80 & 0.3 & 2.727 & $ 0.000^{+2.208}_{-0.000}$ & ${\bf 34.910^{+3.640}_{-3.610}}$ & ${\bf 190.800^{+3.600}_{-3.600}}$ & ${\bf 37.150^{+7.660}_{-7.660}}$ & $0.000^{+1.227}_{-0.000}$ & 3.28 \\
N & $-1$ & 3.4 &  6.80 & 0.3 & 2.727 & $ 6.784^{+9.496}_{-6.784}$ & ${\bf 20.900^{+4.040}_{-4.500}}$ & ${\bf 65.350^{+2.470}_{-2.410}}$ & ${\bf 6.397^{+3.370}_{-3.341}}$ & $0.000^{+5.330}_{-0.000}$ & 0.46 \\
O & $-2$ & 3.4 & 10.20 & 0.4 & 2.727 & ${\bf 10.640^{+1.480}_{-2.033}}$ & $ 0.000^{+0.896}_{-0.000}$ & ${\bf 12.420^{+0.690}_{-0.690}}$ & ${\bf 2.690^{+0.971}_{-1.046}}$ & $1.451^{+2.189}_{-1.451}$ & 0.71 \\
P & $-3$ & 3.4 &  6.80 & 0.3 & 2.857 & ${\bf 22.350^{+2.410}_{-5.220}}$ & $ 0.357^{+2.443}_{-0.357}$ & ${\bf 20.600^{+1.660}_{-1.310}}$ & ${\bf 2.624^{+1.357}_{-1.456}}$ & $0.000^{+2.747}_{-0.000}$ & 0.65 \\
Q & $-4$ & 3.4 &  6.80 & 0.3 & 2.857 & ${\bf 12.450^{+4.090}_{-4.429}}$ & $ 1.277^{+2.065}_{-1.277}$ & ${\bf 16.870^{+1.450}_{-1.440}}$ & $ 1.137^{+1.189}_{-1.137}$ & $0.816^{+2.429}_{-0.816}$ & 0.91 \\
R & $-5$ & 3.4 &  6.80 & 0.3 & 2.857 & ${\bf 12.900^{+1.150}_{-1.510}}$ & $ 0.000^{+0.478}_{-0.000}$ & ${\bf 12.200^{+0.890}_{-0.830}}$ & $ 0.000^{+0.357}_{-0.000}$ & $0.000^{+1.292}_{-0.000}$ & 1.24 

\enddata

\tablenotetext{a}{ \ Same as for Table~\ref{tab:modparams_b}}
\tablenotetext{b}{ \ Same as for Table~\ref{tab:modparams_b}}
\tablenotetext{c}{ \ Derived parameter}
\tablenotetext{d}{ \ Bold values indicate values constrained to a confidence level of 1-sigma.}

\end{deluxetable}


\begin{deluxetable}{ccccccccc}
\rotate
\tablewidth{0pc}
\tablecaption{MASS OF SUBMICRON GRAINS FOR FRAGMENT B \label{tab:minmass_b}}
\tablehead{
 & & \multicolumn{5}{c}{\underbar{Mass Relative to Total Mass}} & & \\
\colhead{Extraction} & \colhead{Total Mass} & \colhead{Amorphous} & \colhead{Amorphous} & \colhead{Amorphous} & \colhead{Crystalline} & \colhead{Ortho-} & \colhead{Silicate/} & \colhead{} \\
\colhead{Aperture} & \colhead{($\times 10^{2}$ kg)} & \colhead{Pyroxene} & \colhead{Olivine} & \colhead{Carbon} & \colhead{Olivine} & \colhead{pyroxene} & \colhead{Carbon} & \colhead{$f_{cryst}$\tablenotemark{a}}
}
\startdata

A & $0.456^{+0.069}_{-0.069}$ & $0.430^{+0.082}_{-0.061}$ & $0.000^{+0.002}_{-0.000}$ & $0.000^{+0.003}_{-0.000}$ & $0.000^{+0.030}_{-0.000}$ & $0.570^{+0.061}_{-0.082}$ & $\cdots$                  & $0.570^{+0.061}_{-0.082}$ \\
B & $2.300^{+0.288}_{-0.282}$ & $0.192^{+0.060}_{-0.059}$ & $0.045^{+0.032}_{-0.028}$ & $0.427^{+0.052}_{-0.041}$ & $0.069^{+0.060}_{-0.069}$ & $0.266^{+0.090}_{-0.113}$ & $1.341^{+0.250}_{-0.253}$ & $0.335^{+0.089}_{-0.112}$ \\
C & $1.999^{+0.099}_{-0.098}$ & $0.309^{+0.048}_{-0.045}$ & $0.000^{+0.004}_{-0.000}$ & $0.464^{+0.014}_{-0.013}$ & $0.031^{+0.034}_{-0.031}$ & $0.197^{+0.052}_{-0.057}$ & $1.156^{+0.063}_{-0.065}$ & $0.228^{+0.051}_{-0.056}$ \\
D & $2.083^{+0.108}_{-0.106}$ & $0.355^{+0.054}_{-0.049}$ & $0.000^{+0.012}_{-0.000}$ & $0.445^{+0.014}_{-0.013}$ & $0.078^{+0.031}_{-0.033}$ & $0.122^{+0.059}_{-0.064}$ & $1.248^{+0.068}_{-0.069}$ & $0.200^{+0.057}_{-0.063}$ \\
E & $1.655^{+0.129}_{-0.129}$ & $0.379^{+0.065}_{-0.059}$ & $0.000^{+0.003}_{-0.000}$ & $0.390^{+0.022}_{-0.019}$ & $0.024^{+0.045}_{-0.024}$ & $0.207^{+0.069}_{-0.081}$ & $1.564^{+0.132}_{-0.137}$ & $0.231^{+0.072}_{-0.084}$ \\
F & $2.090^{+0.099}_{-0.099}$ &  $0.313^{+0.046}_{-0.043}$ & $0.000^{+0.008}_{-0.000}$ & $0.405^{+0.011}_{-0.010}$ & $0.043^{+0.032}_{-0.034}$ & $0.239^{+0.048}_{-0.053}$ & $1.468^{+0.061}_{-0.065}$ & $0.282^{+0.046}_{-0.050}$ \\
G & $1.815^{+0.092}_{-0.093}$ & $0.503^{+0.058}_{-0.054}$ & $0.000^{+0.005}_{-0.000}$ & $0.299^{+0.013}_{-0.014}$ & $0.052^{+0.034}_{-0.036}$ & $0.146^{+0.056}_{-0.062}$ & $2.342^{+0.159}_{-0.137}$ & $0.198^{+0.054}_{-0.060}$ \\
H & $1.456^{+0.083}_{-0.068}$ & $0.713^{+0.067}_{-0.068}$ & $0.000^{+0.004}_{-0.000}$ & $0.226^{+0.017}_{-0.019}$ & $0.000^{+0.038}_{-0.000}$ & $0.061^{+0.068}_{-0.061}$ & $3.430^{+0.402}_{-0.313}$ & $0.061^{+0.068}_{-0.061}$

\enddata
\tablenotetext{a)}{Defined as $f_{cryst}^{silicates} \equiv$ (crystalline)/(crystalline + amorphous).}
\end{deluxetable}

\begin{deluxetable}{ccccccccc}
\rotate
\tablewidth{0pc}
\tablecaption{MASS OF SUBMICRON GRAINS FOR FRAGMENT C \label{tab:minmass_c}}
\tablehead{
 & & \multicolumn{5}{c}{\underbar{Mass Relative to Total Mass}} & & \\
\colhead{Extraction} & \colhead{Total Mass} & \colhead{Amorphous} & \colhead{Amorphous} & \colhead{Amorphous} & \colhead{Crystalline} & \colhead{Ortho-} & \colhead{Silicate/} & \colhead{} \\
\colhead{Aperture} & \colhead{($\times 10^{2}$ kg)} & \colhead{Pyroxene} & \colhead{Olivine} & \colhead{Carbon} & \colhead{Olivine} & \colhead{pyroxene} & \colhead{Carbon} & \colhead{$f_{cryst}$\tablenotemark{a}}
}
\startdata

I &  $0.689^{+0.037}_{-0.021}$ & $0.878^{+0.051}_{-0.081}$ & $0.000^{+0.008}_{-0.000}$ & $0.122^{+0.054}_{-0.051}$ & $0.000^{+0.015}_{-0.000}$ & $0.000^{+0.072}_{-0.000}$ & $7.200^{+5.980}_{-2.503}$ & $0.000^{+0.072}_{-0.000}$ \\
J &  $0.935^{+0.090}_{-0.012}$ & $0.742^{+0.061}_{-0.124}$ & $0.000^{+0.007}_{-0.000}$ & $0.241^{+0.037}_{-0.044}$ & $0.000^{+0.010}_{-0.000}$ & $0.017^{+0.120}_{-0.017}$ & $3.148^{+0.918}_{-0.549}$ & $0.017^{+0.120}_{-0.017}$ \\
K &  $2.174^{+0.113}_{-0.094}$ & $0.541^{+0.082}_{-0.080}$ & $0.000^{+0.007}_{-0.000}$ & $0.391^{+0.026}_{-0.027}$ & $0.000^{+0.012}_{-0.000}$ & $0.069^{+0.075}_{-0.069}$ & $1.559^{+0.187}_{-0.159}$ & $0.069^{+0.075}_{-0.069}$ \\
L &  $1.065^{+0.575}_{-0.573}$ & $0.072^{+0.097}_{-0.029}$ & $0.000^{+0.000}_{-0.000}$ & $0.038^{+0.039}_{-0.012}$ & $0.000^{+0.021}_{-0.000}$ & $0.890^{+0.041}_{-0.136}$ & $25.228^{+11.706}_{-13.266}$ & $0.890^{+0.041}_{-0.136}$ \\
M & $13.240^{+0.640}_{-0.630}$ & $0.000^{+0.009}_{-0.000}$ & $0.144^{+0.018}_{-0.017}$ & $0.598^{+0.029}_{-0.026}$ & $0.257^{+0.039}_{-0.043}$ & $0.000^{+0.008}_{-0.000}$ & $0.671^{+0.076}_{-0.076}$ & $0.257^{+0.039}_{-0.043}$ \\
N &  $4.817^{+0.423}_{-0.365}$ & $0.077^{+0.096}_{-0.077}$ & $0.238^{+0.065}_{-0.065}$ & $0.563^{+0.052}_{-0.053}$ & $0.122^{+0.050}_{-0.059}$ & $0.000^{+0.093}_{-0.000}$ & $0.775^{+0.183}_{-0.150}$ & $0.122^{+0.095}_{-0.059}$ \\
O &  $2.626^{+0.294}_{-0.228}$ & $0.392^{+0.090}_{-0.080}$ & $0.000^{+0.036}_{-0.000}$ & $0.347^{+0.034}_{-0.024}$ & $0.169^{+0.047}_{-0.056}$ & $0.091^{+0.115}_{-0.091}$ & $1.882^{+0.216}_{-0.259}$ & $0.261^{+0.104}_{-0.098}$ \\ 
P &  $2.921^{+0.184}_{-0.184}$ & $0.535^{+0.047}_{-0.104}$ & $0.009^{+0.063}_{-0.009}$ & $0.374^{+0.045}_{-0.031}$ & $0.082^{+0.038}_{-0.044}$ & $0.000^{+0.081}_{-0.000}$ & $1.675^{+0.245}_{-0.287}$ & $0.082^{+0.083}_{-0.044}$ \\
Q &  $2.034^{+0.173}_{-0.186}$ & $0.428^{+0.124}_{-0.138}$ & $0.044^{+0.083}_{-0.044}$ & $0.440^{+0.061}_{-0.048}$ & $0.051^{+0.046}_{-0.051}$ & $0.037^{+0.098}_{-0.037}$ & $1.275^{+0.277}_{-0.279}$ & $0.088^{+0.105}_{-0.073}$ \\
R &  $1.549^{+0.069}_{-0.027}$ & $0.582^{+0.037}_{-0.070}$ & $0.000^{+0.022}_{-0.000}$ & $0.418^{+0.031}_{-0.037}$ & $0.000^{+0.021}_{-0.000}$ & $0.000^{+0.073}_{-0.000}$ & $1.395^{+0.232}_{-0.166}$ & $0.000^{+0.073}_{-0.000}$

\enddata
\tablenotetext{a)}{Defined as $f_{cryst}^{silicates} \equiv$ (crystalline)/(crystalline + amorphous).}
\end{deluxetable}

\clearpage


\end{document}